\documentclass{agujournal2019}
\usepackage{url} 
\usepackage[inline]{trackchanges} 
\usepackage{soul}

\usepackage[english]{babel}
\usepackage{amssymb}
\usepackage{amsmath}

\journalname{Journal of Advances in Modeling Earth Systems (JAMES)}

\newcommand{\ansA}[1]{{\color{black}#1}}

\begin{document}

%
%

\title{On energy-aware hybrid models}

%
%




\authors{Igor Shevchenko and Dan Crisan}


\affiliation{1}{National Oceanography Centre, European Way, Southampton, SO14 3ZH, UK}
\affiliation{2}{Department of Mathematics, Imperial College London,180 Queen's Gate, London, SW7 2AZ, UK}




\correspondingauthor{Igor Shevchenko}{igor.shevchenko@noc.ac.uk}



\begin{keypoints}
\item Deterministic and stochastic energy-aware hybrid models have been proposed
\item Hybrid models allow low-resolution simulations without compromising on quality compared with high-resolution runs
\item Hybrid approach produce more accurate ensemble predictions than their classical GFD counterparts
\end{keypoints}

%
%

%
%


\begin{abstract}
This study proposes deterministic and stochastic energy-aware hybrid models that \ansA{should} enable simulations of \ansA{idealized} and \ansA{primitive-equations} Geophysical Fluid Dynamics (GFD) models at low resolutions without compromising on quality compared with high-resolution runs. Such hybrid models bridge the data-driven and physics-driven modelling paradigms by combining regional stability and classical GFD models at low resolution that cannot reproduce high-resolution reference flow features (large-scale flows and small-scale vortices) which are, however, resolved. Hybrid models use an energy-aware correction of advection velocity and extra forcing compensating for the drift of the low-resolution model away from the reference phase space. The main advantages of hybrid models are that they allow for physics-driven flow recombination within the reference energy band, reproduce resolved reference flow features, and produce more accurate ensemble forecasts than their classical GFD counterparts.

Hybrid models offer appealing benefits and flexibility to the modelling and forecasting communities, as they are computationally cheap and can use both numerically-computed flows and observations from different sources. All these suggest that the hybrid approach has the potential to exploit low-resolution models for long-term weather forecasts and climate projections thus offering a new cost effective way of GFD modelling.

The proposed hybrid approach has been tested on a three-layer quasi-geostrophic model for a beta-plane Gulf Stream flow configuration. The results show that the low-resolution hybrid model reproduces the reference flow features that are resolved on the coarse grid and also gives a more accurate ensemble forecast than the physics-driven~model.
\end{abstract}

\section*{Plain Language Summary}
Reliable weather forecast and climate prediction are crucial for socio-economic sectors, decision making, and strategic planning for mitigating risks and impacts of natural disasters. In order to forecast weather and climate, observations and numerical models are used. A necessary ingredient for these forecasts is the ocean-atmospheric model that resolves mesoscale oceanic eddies and can use large ensembles (hundred of members). Supercomputers can run eddy-resolving simulations but only for single runs or use large ensembles but in non-eddy-resolving regimes. Applying the large ensemble approach with eddy-resolving simulations is far beyond what supercomputers will be able to compute for the next decades, while it is urgently needed.

This study offers deterministic and stochastic energy-aware hybrid models that enable simulations of Geophysical Fluid Dynamics models at low resolutions without compromising on quality compared with high-resolution runs. The use of low resolutions makes hybrid models much faster than high-resolution physics-driven runs. This acceleration can be translated into the use of larger ensembles. Thus, the proposed hybrid approach has the potential to exploit large ensembles of high-quality solutions for long-term weather forecasts and climate projections for the very first time.

%
%

%


%
%
%
%

\section{Introduction} 
The surge in data-driven modelling has currently influenced the traditional course of Geophysical Fluid Dynamics and split it into two branches. One branch continues to use the physics-driven approach based on
classical  models (e.g.~\cite{Vallis2016}) and parameterizations
(e.g.~\cite{GentMcwilliams1990,DuanNadiga2007,Frederiksen_et_al2012,
PortaMana_Zanna2014,CooperZanna2015,
Grooms_et_al2015,Berloff_2015,Berloff_2016,Berner_et_al2017,Berloff_2018,Ryzhov_etal_2019,
CCHWS2019_1,Ryzhov_etal_2020,CCHWS2019_3,CCHWS2020_4,CCHPS2020_J2,Kitsios_et_al2023}), while
the other started employing the data-driven approach (e.g.~\cite{Agarwal_et_al2021}).
The physics-driven approach is believed to provide more accurate than current future projections at resolutions only achievable on at least exascale computers. Although the era of exascale computations is picking up speed in the world, routine long-time
exa-simulations and their analysis is still to be achieved, while the need in reliable climate models is urgent given the current climate concerns.
On the other hand, data-driven approaches promise faster simulations,
while providing a similar level of fidelity to the physics-driven models. Currently, many data-driven approaches
exploit artificial neural networks (e.g.~\cite{Dramsch2020,BNK_2022}) trained on reference runs of physics-driven models.

In this work we offer a hybrid approach that combines classical models and the data-driven hyper-parameterization (HP) approach  (e.g.~\cite{SB2021_J1,SB2022_J2,SB2022_J1,SB2022_J3}). 
The later shifts the focus from the physical space to the phase space also known as state space. It considers the inability of the low-resolution model 
to reproduce resolved reference flow structures as the persistent tendency of the phase space trajectory (the low-resolution solution) to escape the reference phase space (the phase space of the reference solution);
{\it note that the term ``low-resolution'' in the context of HP approach is relative, and only reflects the fact
that the resolution of the HP solution is lower than that of the reference solution}.
The philosophy behind the hybrid approach is to augment the classical model at low resolution with reference data in order to ensure that a given measure of goodness holds. 
In this work, the measure of goodness is that the solution of the hybrid model (or hybrid for short) lies within the reference energy band. 
It is chosen so that to ensure a long-term evolution of the hybrid on the reference energy manifold and to retain reference flow features that are resolved at low resolution; the reference energy manifold is a hyper-surface (in the phase space) on which the energy of the reference solution lies within a prescribed energy band.
In a nutshell, the hybrid approach rests on the hypothesis that nonlinear interactions in the hybrid model will improve
intra- and inter-scale energy transfers towards the reference ones through the control of energy on specific scales.

The strong point of the hybrid approach is that it allows for any physics-driven flow recombination within the reference energy band and reproduces resolved reference flow features.
\ansA{A physics-driven flow recombination within the reference energy band means hybrid model solutions that exist on the reference energy manifold.
In other words, the hybrid solution is not restricted to evolve within a neighborhood of the reference trajectory in an attempt to predict it beyond the sampled interval (as in conventional methods), but to evolve in a much larger phase space corresponding to the reference energy manifold.}

The hybrid approach offers a wide range of advantages and a great flexibility to ocean and atmosphere modellers working with models and observations. Namely, (1) hybrid models
are computationally inexpensive (a hybrid run is on the scale of the coarse-grid model simulation); (2) they can use both numerically-computed flows and measurements from different sources (drifters, weather stations, satellites, etc.) as reference data; 
and (3) require minimalistic interventions into the dynamical core of the underlying model.

\section{Hybrid hyper-parameterization models}
The hybrid approach assumes that a low-resolution model which cannot reproduce high-resolution reference flow features 
(which are, however, resolved at low resolution) can be turned into one that is capable of doing so. It
is possible to achieve this if (1) the total energy of the hybrid model lies within the reference energy band and 
(2) the hybrid model respects the regional stability property.  
We define the latter as a property of the solution of a differential equation\\
\begin{equation}
\frac{d\mathbf{x}}{dt}=\mathbf{F}(\mathbf{x}),\quad \mathbf{x}\in\mathbb{R}^n
\label{eq:ode1} 
\end{equation}
to permanently \ansA{remain in a neighborhood of the phase space region} if the initial condition is 
sufficiently close to that region.

The first hybrid HP model has been proposed in~\cite{SB2022_J2}. Its dynamical core comprises a quasi-geostrophic (QG) model augmented with
the constrained optimization machinery that confines the flow dynamics to evolve within the reference phase space. 
The hybrid approach proposed in this study is another step forward towards the use of low-resolution GFD models in long-term climate projections.
At the core of the approach lies hybrid models using the following:\\ 
\indent $\bullet$ an advection velocity corrector $\mathbf{A}$;\\
\indent $\bullet$ an extra forcing $\mathbf{G}$ compensating the drift of the model away from the reference phase space;\\
\indent $\bullet$ a multi-scale decomposition $\mathbf{M}$;\\ 
\indent $\bullet$ an optimization method that searches for spatial scale amplitudes such that the total energy of the hybrid is within the reference band.

For the illustration purposes, we consider the transport equation for a quantity~$\phi$\\
\begin{equation}
\partial_t\phi+\mathbf{v}\cdot\nabla\phi=\mathbf{F}(\phi)
\label{eq:pde} 
\end{equation}
with $\mathbf{v}(t,\cdot)$ being the velocity vector; the dot in the argument of $\mathbf{v}$ means the space dependence.
Typically, \eqref{eq:pde} is solved at high resolution, and every run is computationally too intensive to allow for long-term or large ensemble simulations.
An affordable option would be a hybrid model that could run at lower resolution on the one hand, while 
reproducing resolved on the coarse-grid flow features of the reference high-resolution flow dynamics on the other.

We propose the following hybrid model corresponding to~\eqref{eq:pde}:\\
\begin{equation}
\partial_t\psi+(\mathbf{u}+\mathbf{A}(\mathbf{u},\mathbf{v}))\cdot\nabla\psi=\mathbf{F}(\psi)+\mathbf{G}(\psi,\phi),
\quad \mathbf{G}(\psi,\phi):=\eta(\mathbf{M}(\psi,\phi)-\psi)
\label{eq:hybrid} 
\end{equation}
with $\eta$ being the nudging strength, \ansA{$\mathbf{u}$ is the velocity vector}, and $\mathbf{M}$ is the multi-scale decompositions defined as\\
\begin{equation}
\mathbf{M}(\psi,\phi):=\sum\nolimits^S_{s=1}\lambda_s\mathcal{M}_s(\psi,\widehat{\phi}),\quad 
\widehat{\phi}:=\frac{1}{M}\sum\nolimits_{i=\mathcal{U}_I}\mathcal{P}(\phi_i)\Big|_{\mathcal{U}(\psi(t,\cdot))},
\label{eq:m} 
\end{equation}
where $\mathcal{M}_s  : \widehat{\phi}\rightarrow\widehat{\phi}_s$ is an operator extracting the $s$-scale flow dynamics, $\widehat{\phi}_s$, from $\widehat{\phi}$, which includes all 
(both underresolved and resolved on the computational grid) spatial scales,
$\lambda_s$ is the amplitude of $\widehat{\phi}_s$, $S$ is the total number of scales in the decomposition,
$\mathcal{P}$ is a projector from high to low resolution,
$\mathcal{U}(\psi(t,\cdot))$ is a neighbourhood of $\psi(t,\cdot)$ in the reference phase space.
\ansA{This neighborhood is a set of $M$ fields $\phi_i$ and $\mathbf{v}_i$ ($i\in\mathcal{U}_I$) nearest in $l_2$-norm to the hybrid solution $\psi(t,\cdot)$;
$\mathcal{U}_I$ is a set of timesteps indexing the discrete reference solutions 
$\phi$ and $\mathbf{v}$ in the neighbourhood of $\psi(t,\cdot)$.
Equation~\eqref{eq:hybrid} is solved on a grid coarser than the one used in equation~\eqref{eq:pde}, and $\phi$ needs, therefore, to be projected from the fine to the coarse grid. It is worth mentioning again
that $\mathbf{u}$ is the flow velocity in equation~\eqref{eq:pde}, and $\mathbf{v}$ is the flow velocity 
in the hybrid model~\eqref{eq:hybrid}; both $\mathbf{u}$ and $\mathbf{v}$ are velocities at the coarse grid in what follows.
}

The velocity corrector\\
\begin{equation}
\mathbf{A}(\mathbf{u},\mathbf{v}):=\sum\nolimits^S_{s=1}\gamma_s\mathcal{M}_s(\widehat{\mathbf{v}}),\quad 
\widehat{\mathbf{v}}:=\frac{1}{M}\sum\nolimits_{i=\mathcal{U}_I}\mathcal{P}(\mathbf{v}_i)\Big|_{\mathcal{U}(\psi(t,\cdot))}
\label{eq:a} 
\end{equation}
with index $i$ being the same as the one in~\eqref{eq:m},
\ansA{and $\gamma_s$ is the amplitude of $\widehat{\mathbf{v}}_s$; note that the CFL condition needs to be modified to account for 
$\mathbf{A}$.}
An alternative to the velocity corrector is to consider the following stochastic correction\\
\begin{equation}
\mathbf{A}(\mathbf{u},\mathbf{v}):=\sum\nolimits^S_{s=1}\gamma_s\mathcal{M}_s(\widehat{\mathbf{v}})dt+\sum\nolimits^S_{s=1}\mathcal{M}_s(\widehat{\mathbf{v}})\circ dW^s_t,
\label{eq:dw} 
\end{equation}
with independent Brownian motions $dW^s_t$. In this case the hybrid model~\eqref{eq:hybrid} becomes:\\
\begin{equation}
d\psi+(\mathbf{u}\,dt+\mathbf{A}(\mathbf{u},\mathbf{v}))\cdot\nabla\psi=(\mathbf{F}(\psi)+\mathbf{G}(\psi,\phi))\,dt\,.
\label{eq:shybrid} 
\end{equation}
The velocity correction~\eqref{eq:dw} turns the hybrid model~\eqref{eq:shybrid} into a stochastic hybrid, which is a SALT-class model~\cite{holm2015variational}.
The principle difference with the original SALT approach is that instead of the multi-scale decomposition it uses Empirical Orthogonal Functions which are calibrated on the reference data before the simulation (e.g.~\cite{CCHWS2019_1,CCHWS2019_3,CCHWS2020_4,CCHPS2020_J2,CLLLR2023}). 
In the hybrid approach, the calibration is performed on the fly (by searching for optimal amplitudes) as the simulation goes on.
We will consider both the deterministic and stochastic versions of velocity correction and study how they influence the flow dynamics.

\ansA{
It is important to explain that
the hybrid model looks at the reference data set as being a cloud of points which are not ordered in time (figure~\ref{fig:ref_space}a); 
it is why we write $\phi_i$ and 
$\mathbf{v}_i$ instead of $\phi(t)$ and $\mathbf{v}(t)$. At every time step, the hybrid model takes $M$ nearest $\phi_i$ and $\mathbf{v}_i$ in $l_2$-norm to its solution $\psi(t,\cdot)$ points, see figure~\ref{fig:ref_space}b.
Thus, the hybrid model never runs out of reference data, because one can always find $M$ nearest points to the hybrid solution $\psi(t,\cdot)$.\\
\begin{figure}[h]
\hspace*{-0.5cm}
\begin{tabular}{cc}
(a) & (b)\\
\begin{minipage}{0.5\textwidth}\includegraphics[scale=1.5]{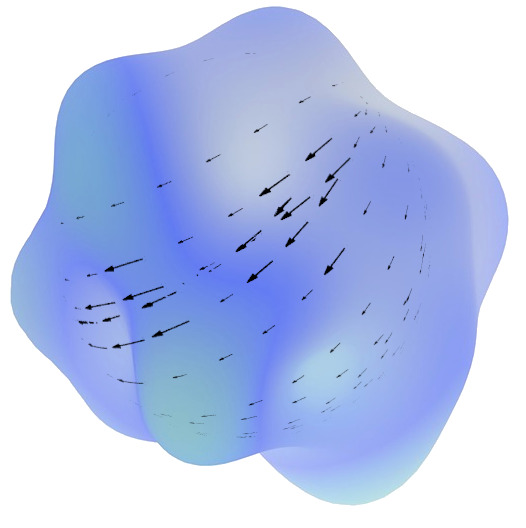}\end{minipage} &
\begin{minipage}{0.5\textwidth}\includegraphics[scale=1.5]{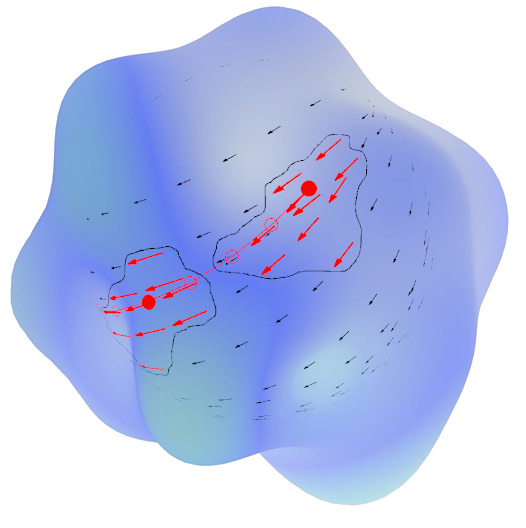}\end{minipage}\\
\end{tabular}
\caption{\ansA{Shown is  a sketch of (a) the reference phase space (blue blob), vectors $\{\mathbf{v}_{i\in[1,N]}\}$ (black arrows) pointing out from points 
$\{\phi_{i\in[1,N]}\}$ (from the hybrid model point of view all these points forms a cloud of points which are not ordered in time); (b) the reference phase space with the hybrid solution $\psi$ (red curve) and 
the neighborhood $\mathcal{U}(\psi(t,\cdot))$ shown as a black contour surrounding a red dot, which is the hybrid solution $\psi(t,\cdot)$
at time $t$. The red vectors (and also every point they point out from) are those $M$ fields $\mathbf{v}_i$ (and $\phi_i$), $i\in\mathcal{U}_I$, nearest in $l_2$-norm to the hybrid solution $\psi(t,\cdot)$.
}}
\label{fig:ref_space}
\end{figure}
}

{\bf Reference data acquisition}. In order to get the reference solution 
$\{\phi_{i\in[1,N]}\}$ and velocity $\{\mathbf{v}_{i\in[1,N]}\}$, with $N$ being the number of records, for the hybrid HP model~\eqref{eq:hybrid}, we simulate the reference model~\eqref{eq:pde} at high resolution and project $\phi$ and $\mathbf{v}$ onto the grid used in the hybrid model; note that the high-resolution model runs only once (before the hybrid simulation) to acquire reference data, this reference run is assumed to be considerably shorter \ansA{(in terms of reference records)} than that of the hybrid model.
If also observations are available, one can project them onto the hybrid model grid and then add them to the reference data set.
In this case the reference data set becomes $\{\mathbf{\phi}_{i\in[1,N]},\widetilde{\mathbf{\phi}}_{j\in[1,N_o]}\}$ and
$\{\mathbf{v}_{i\in[1,N]},\widetilde{\mathbf{v}}_{j\in[1,N_o]}\}$, where the tilde stands for the observed variable,
\ansA{and $N_o$ is the number of observational records}.
In case of no numerical simulations available, one can solely use observations.

\ansA{
It is important to reiterate that the hybrid model~\eqref{eq:hybrid}  requires the $\phi$ and $\mathbf{v}$ fields at each grid point in order to 
be consistent with its spatial discretization. It is a very realistic and an easy-to-achieve requirement,
as realistic ocean-atmospheric simulations and observational products now have 1/12$^\circ$ resolution and higher.
In our study, we project point-to-point the high-resolution solution onto the coarse grid implicitly assuming that the coarse grid is a subgird of
the high-resolution grid. If this is not the case one can use interpolation schemes, spectral filters, spatial averaging the fine-grid solution over the coarse grid cell, etc. The point-to-point projection between two grids we use is a bijection between the nodes (with the same indices) of the grids.
The reference data is not required to be sampled every time step used in the hybrid model to be consistent 
with its integration scheme, because the reference data do not depend on time from the hybrid model point of view.}

{\bf The multi-scale decomposition} controls the amount of energy added to or removed from the hybrid model on a given scale.
Operator $\mathcal{M}_s$ can be thought of as a spatial filter or an evolutionary model computing the s-scale flow $\widehat{\phi}_s$. 
We use the spectral filtering (e.g.~\cite{Sagaut2006}) \ansA{based on the discrete Fourier transform to extract flow dynamics for specific spatial scales. It is a three-step process: (1) Fourier transform the spatial field;
(2) set to zero Fourier harmonics with the wave numbers which do not satisfy the chosen scales;
(3) inverse Fourier transform back to the original space domain.}
The choice is not limited to the spectral filtering, and can be chosen based on the problem-dependent requirements. 

{\bf The neighborhood} $\mathcal{U}(\psi(t,\cdot))$ \ansA{is a set of $M$ fields $\phi_i$ and $\mathbf{v}_i$ ($i\in\mathcal{U}_I$) nearest in $l_2$-norm to the hybrid solution $\psi(t,\cdot)$ in the reference phase space}, i.e. in the phase space of model~\eqref{eq:pde}.
For all simulations presented below, we take $M=10$ (as in the first HP method~\cite{SB2021_J1}) and note that this parameter is of minor importance, 
as the optimization method delivers amplitudes $\lambda_s$ 
and $\gamma_s$ to ensure the energy of the hybrid model is within the reference energy band.
\ansA{The choice of the norm and the way the neighborhood $\mathcal{U}(\psi(t,\cdot))$ is calculated is not limited to those used in this study,
and can vary depending on what is needed. Our choice is, probably, the simplest one, but it is enough for the purpose of this work.}

{\bf The number of spatial scales} $S$ is calculated based on what the hybrid model can resolve and what it can not, \ansA{and how much the reference and the low-resolution energies differ at certain scales.
We deem a scale to be resolved if there are 10 grid points per scale;
it is associated with the dispersion error of the CABARET scheme~\cite{Karabasov_et_al2009} which is used to compute the QG model.
}

\ansA{
In order to calculate the optimal multi-scale decomposition minimizing
the difference between the reference solution and the hybrid solution over the sampled interval,
one should run a computationally very intensive optimization procedure which is beyond the scope of this paper. 
This is hardly of any practical value, as it is limited to the QG setup only used in the study. 
More importantly, this optimal decomposition is unnecessary, since our measure of goodness, is for the hybrid solution to lie within the reference energy band, is satisfied.

In order to calculate the multi-scale decomposition {\it which satisfies our measure of goodness}
we analyse the energy at different scales for the reference and low-resolution solutions.
If this difference is relatively large (compared to other scales) for certain scales then these scales are selected for the decomposition.
We use the spectral filtering in the scale decomposition, therefore it is natural to use the energy spectral density, $E_S$, instead of the kinetic and potential energies, as it would require the recalculation of these energies on the scales used in the decomposition.

Getting a bit ahead, we consider a two (figure~\ref{fig:scales2_3}a) and three (figure~\ref{fig:scales2_3}b) scale decompositions within 
the context of QG model (we consider the QG model in section~\ref{sec:qg}).
First, we look at the 3-scale decomposition (figure~\ref{fig:scales2_3}b). 
The $E_S$ of the reference solution (blue bar) and of the low-resolution solution (red bar)
has a much larger difference on the scales $[30,300)\, {\rm km}$ and $[300,1900)\, {\rm km}$ compared to the one on
the scales $[1900,3840]\, {\rm km}$. It means that the energy on the scales $[1900,3840]\, {\rm km}$ in the low-resolution model 
is very close to the reference energy, and making this difference even smaller (by including these scales in the optimization described below) is unnecessary, 
as the contribution of this difference in the total energy of the low-resolution model is much smaller compared to that of the scales $[30,300)\, {\rm km}$ and $[300,1900)\, {\rm km}$. Thus, considering the scales $[1900,3840]\, {\rm km}$ in the multi-scale decomposition is unnecessary. Therefore, we combine the scales $[300,1900)\, {\rm km}$ and 
$[1900,3840]\, {\rm km}$ into the scales $[300,3840]\, {\rm km}$ and use a 2-scale decomposition (figure~\ref{fig:scales2_3}a).\\
\begin{figure}[h]
\hspace*{-0.75cm}
\begin{tabular}{cc}
\hspace*{-5.75cm} (a) & \hspace*{-5.75cm} (b)\\
\begin{minipage}{0.5\textwidth}\includegraphics[scale=0.125]{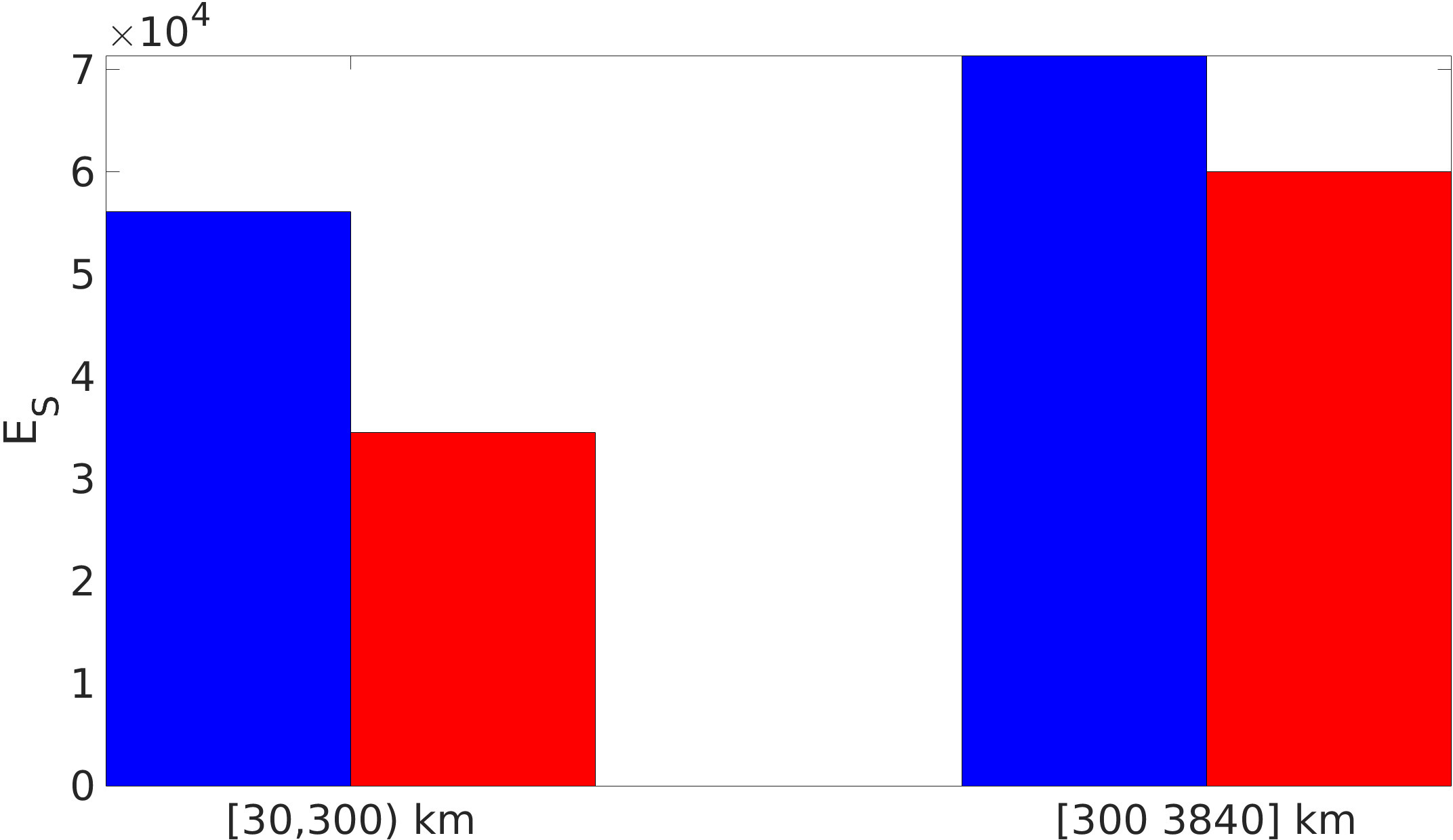}\end{minipage} &
\begin{minipage}{0.5\textwidth}\includegraphics[scale=0.125]{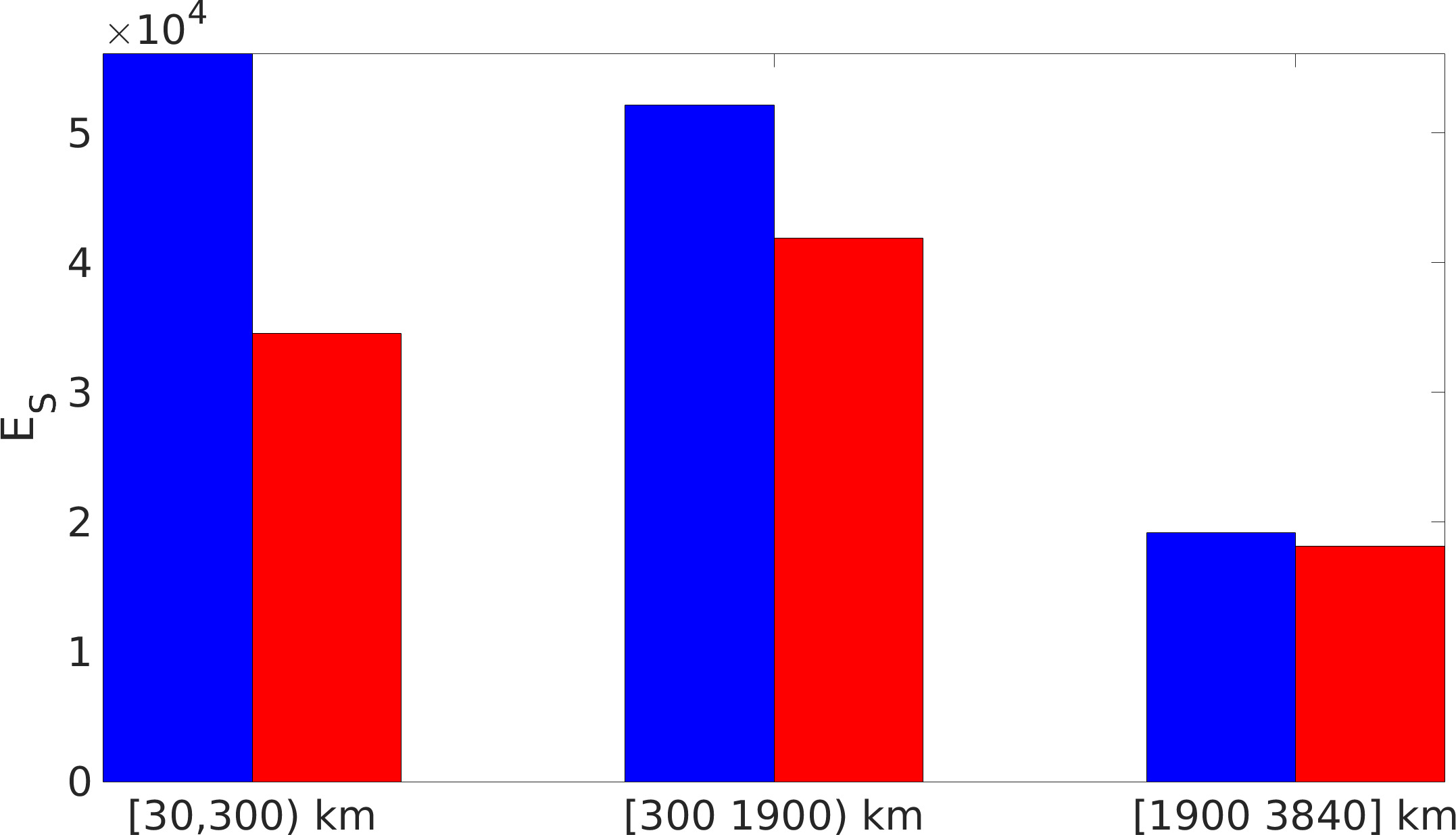}\end{minipage}\\
\end{tabular}
\caption{\ansA{Shown is the energy spectral density, $E_s$, for a two (a) and three (b) scale decomposition for the reference 
solution (blue) and the low-resolution solution (red) of the QG model~\eqref{eq:pve}. The $E_s$ is computed for the time-mean potential
vorticity anomaly $\mathbf{q}$ over the 1-year long reference period.
}}
\label{fig:scales2_3}
\end{figure}
}

Thus, we use a 2-scale decomposition ($S=2$), which splits the flow into two: the first one includes all underresolved scales, while the second includes all resolved scales. A more general approach would be to compare the total energy of the reference and low-resolution solutions,
and decompose the flow into scales the energy on which needs correction.

{\bf Energy calculation}. In order to keep the total energy of the hybrid model within the reference energy band, we compute kinetic and potential energies. 
If potential energy is not available then one should use only the kinetic energy which is typically a standard diagnostic in ocean and atmospheric models. Note that it is more efficient
to calculate the energies for all reference data before running the hybrid model, as it allows to avoid unnecessary recomputations of energy during the simulation.

{\bf Optimization method}.
The last ingredient of the hybrid approach is the optimization method.
We use Powell's method~\cite{Powell1964} which is a derivative free optimization method, and avoid methods computing gradients, as it is an unnecessary complication for this study. 
Note that the nudging strength $\eta$ can also be included in the optimization, but we keep it constant 
at $\eta=0.02$ \ansA{(it is a non-dimensional value, the dimensional value of the relaxation time for the nudging parameter
is $\sim$1736 days)} in this work and optimize for scale amplitudes $\lambda_s$ and $\gamma_s$ such that for $\forall t\ge0$ one of the following criteria holds:\\
\begin{equation}
 \begin{aligned}
(C1) & & \|\overline{E}(\phi)-E(\psi(t,\cdot))\|_2\le \varepsilon  \\
 \end{aligned}
\label{eq:opt1}
\end{equation}
\begin{equation}
 \begin{aligned}
(C2) & & \|E(\widehat{\phi})-E(\psi(t,\cdot))\|_2\le \varepsilon
\end{aligned}
\label{eq:opt2}
\end{equation}
\ansA{where $E(\psi(t,\cdot))$ is the total energy of the hybrid solution $\psi(t,\cdot)$, 
$\overline{E}(\phi)$ is the time mean total energy of the reference solution $\phi$ (the coarse-grained high resolution solution), $E(\widehat{\phi})$ is the total energy of $\widehat{\phi}$, and $\varepsilon$ is a tolerance (in this study $\varepsilon=10^{-6}$). 
The goal of the optimization method is to match the reference energy 
$\overline{E}(\phi)$ or $E(\widehat{\phi})$ (depending on the criterion used) to the energy of the hybrid
solution $E(\psi(t,\cdot))$.
}

The optimization method can significantly affect not only the accuracy but also the performance of the hybrid approach. 
For example, the orbital search optimizes for amplitudes to ensure that starting from an initial condition $\psi(t_i,\cdot)$
one of the optimization criteria is satisfied for $\psi(t_j,\cdot)$, $j\gg i$. 
\ansA{In other words, in every iteration of the optimization procedure the model needs to be run over the period $[t_i,t_j]$
to find optimal amplitudes. The orbital search does not optimize at every time step within the interval $[t_i,t_j]$.
Instead, it steers the whole trajectory over $[t_i,t_j]$ by optimizing for only one set of amplitudes.}
Thus, the orbital search along long trajectories can take a considerable toll on the \ansA{efficiency of the simulation}.
Therefore, one should find the right balance between performance and accuracy for a given problem. 
We optimize in the last time step every 24 hours; the time step is 0.5 hour.
As an alternative, one can engage the orbital search over the 24-hour trajectory;
\ansA{the orbital search over a 24-hour trajectory optimizes for amplitudes to ensure that starting 
from an initial condition $\psi(t_i,\cdot)$ one of the optimization criteria holds for $\psi(t_j,\cdot)$, where $t_j-t_i=24$ hours.}
We have tried the latter too, but did not observe any improvement compared to the former.

\section{Multilayer quasi-geostrophic model\label{sec:qg}}
In this section we apply the hybrid approach to the three-layer quasi-geostrophic (QG) model for the evolution of potential vorticity (PV) anomaly $\mathbf{q}=(q_1,q_2,q_3)$~\cite{Pedlosky1987}:\\
\begin{equation}
\partial_t q_j+\mathbf{v}_j\cdot\nabla q_j=F,\quad F:=\delta_{1j}F_{\rm w}-\delta_{j3}\,\mu\nabla^2\psi_j+\nu\nabla^4\psi_j-\beta\psi_{jx}, \quad j=1,2,3\, ,
\label{eq:pve}
\end{equation}
with $\boldsymbol{\psi}=(\psi_1,\psi_2,\psi_3)$ being the velocity streamfunction, $\delta_{ij}$ is the Kronecker symbol, 
and $\mathbf{v}$ is the velocity vector; the planetary vorticity gradient is $\beta=2\times10^{-11}\, {\rm m^{-1}\, s^{-1}}$,
the bottom friction is $\mu=4\times10^{-8}\, {\rm s^{-1}}$, and the lateral eddy viscosity is $\nu=50\, {\rm m^2\, s^{-1}}$.
The asymmetric wind curl forcing, driving the double-gyre ocean circulation, is given~by\\
\[
\displaystyle
F_{\rm w}=\left\{ \begin{array}{ll}
\displaystyle
-1.80\,\ansA{A}\,\tau_o\sin\left(\pi y/y_0\right), & y\in[0,y_0), \\
\displaystyle
{\color{white}-}2.22\,\ansA{A}\,\tau_o\sin\left(\pi (y-y_0)/(L-y_0)\right), & y\in[y_0,L],\\
\end{array}\right.
\]
with the wind stress amplitude $\tau_0=0.03\, {\rm N\, m^{-2}}$ and the tilted zero forcing line
$y_0=0.4L+0.2x$, $x\in[0,L]$; $\ansA{A=\pi/(L\rho_1H_1)}$, $\rho_1=1000\,\rm kg\cdot m^{-3}$ is the top layer density,
and $H_1=250\, \rm m$ is the top layer depth.
The computational domain $\Omega=[0,L]\times[0,L]\times[0,H]$ is a closed, flat-bottom basin with $L=3840\, \rm km$, and the total depth
$H=H_1+H_2+H_3$ given by the isopycnal fluid layers of depths (top to bottom): 
$H_1=0.25\, \rm km$, $H_2=0.75\, \rm km$, $H_3=3.0\, \rm km$. 

The PV anomaly $\boldsymbol{q}$ and the velocity streamfunction $\boldsymbol{\psi}$ are coupled through the system of elliptic equations:\\
\begin{equation}
\boldsymbol{q}=\nabla^2\boldsymbol{\psi}-{\bf S}\boldsymbol{\psi} \, ,
\label{eq:pv}
\end{equation}
with the stratification matrix\\
\[
{\bf S}=\left(\begin{array}{lll}
   {\color{white}-}1.19\cdot10^{-3} & -1.19\cdot10^{-3}                  &  {\color{white}-}0.0                  \\
  -3.95\cdot10^{-4}                 &  {\color{white}-}1.14\cdot10^{-3}  & -7.47\cdot10^{-4}                  \\
          {\color{white}-}0.0          & -1.87\cdot10^{-4}                  &  {\color{white}-}1.87\cdot 10^{-4} \\
\end{array}\right).
\]
\noindent
The stratification parameters are given in units of $\rm km^{-2}$ and chosen so that the first and second Rossby deformation 
radii are $Rd_1=40\, {\rm km}$ and $Rd_2=23\, {\rm km}$, respectively; the choice of these parameters is typical for the North Atlantic,
as it allows to simulate a more realistic (than in different QG setups) but yet idealized eastward jet extension of the western boundary currents.

System~(\ref{eq:pve})-(\ref{eq:pv}) is augmented with the mass conservation constraint~\cite{McWilliams1977}:\\
\begin{equation}
\partial_t\iint\nolimits_{\Omega}(\psi_j-\psi_{j+1})\ dydx=0,\quad j=1,2
\label{eq:masscon}
\end{equation}
with a zero initial condition and a partial-slip boundary condition~\cite{HMG_1992}:\\
\begin{equation}
\left(\partial_{\bf nn}\boldsymbol{\psi}-\alpha^{-1}\partial_{\bf n}\boldsymbol{\psi}\right)\Big|_{\partial\Omega}=0 \, ,
\label{eq:bc}
\end{equation}
where $\alpha=120\, {\rm km}$ is the partial-slip parameter, and $\bf n$ is the normal-to-wall unit vector.

\begin{figure}
\hspace*{-0.75cm}
\begin{tabular}{ll}
\hspace*{0.4cm}(a)  & \hspace*{0.4cm}(b) \\
\begin{minipage}{0.5\textwidth}\includegraphics[scale=0.13]{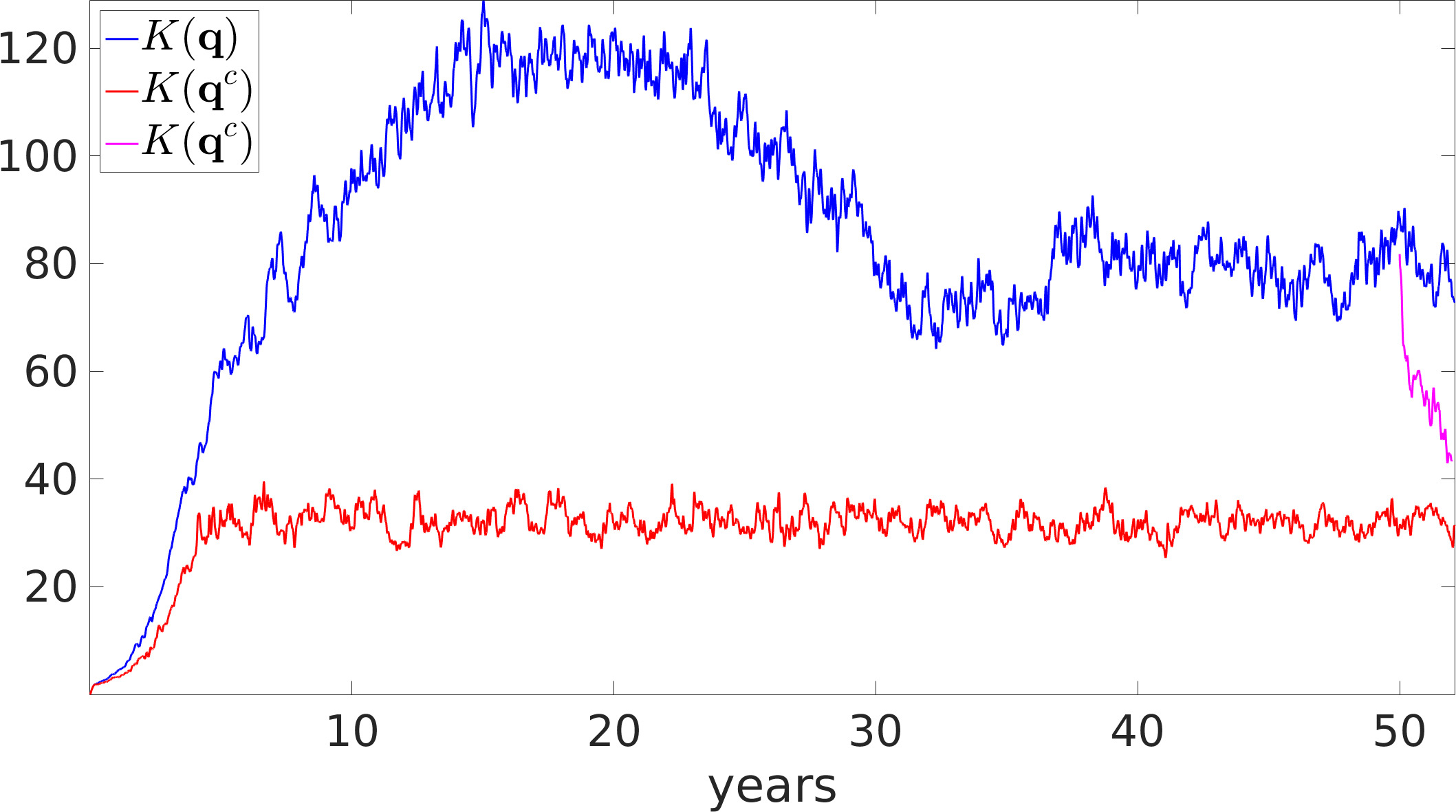}\end{minipage} &
\begin{minipage}{0.5\textwidth}\includegraphics[scale=0.13]{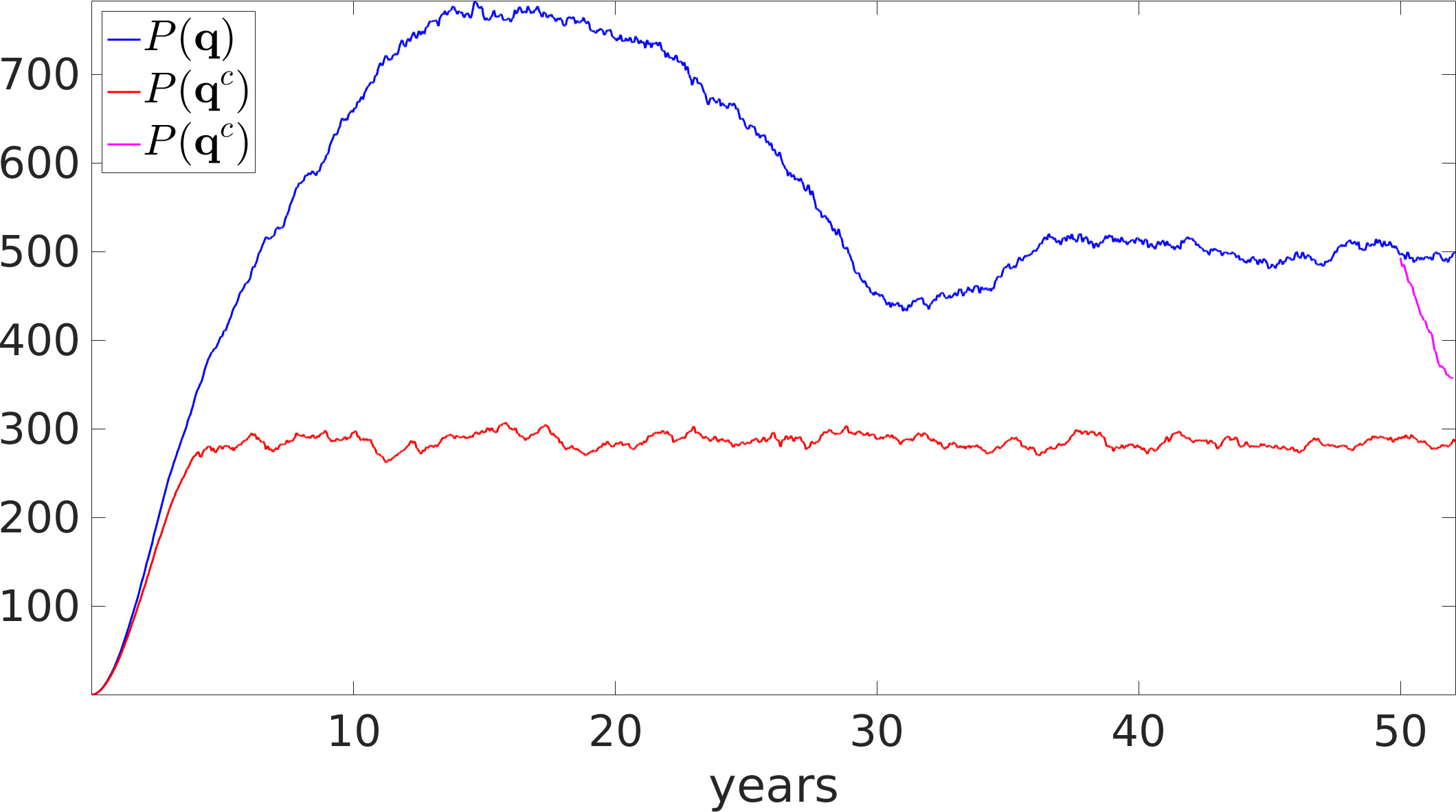}\end{minipage}\\
\end{tabular}
\caption{Time series of the non-dimensional (a) kinetic and (b) potential energies for the reference $\mathbf{q}$ (blue) and low-resolution $\mathbf{q^c}$ (red) solutions over the interval of 52 years. We also show the energies of the low-resolution solution started from the reference initial condition after the spinup (magenta). These energies undergo a rapid and significant drop compared with the reference energy level.}
\label{fig:energy}
\end{figure}

Before proceeding to the hybrid QG model, we explain how to compute the reference data $\{\mathbf{q}_i,\mathbf{v}_i\}_{i\in[1,N]}$.
In order to generate it, we first spin up the QG model for 50 years on a high-resolution 
grid of size $513\times513$ (grid step is $dx=dy=7.5\, {\rm km}$) until the solution is statistically-equilibrated on the energy manifold (figure~\ref{fig:energy}), we then
run the model for another 2 years, and project point-to-point the 2-year solution onto the coarse grid of size $129\times129$ (grid step is $dx=dy=30\, {\rm km}$). Finally, we take only the first year as the reference solution, and keep the second year for validation of the hybrid model.

\ansA{The potential and kinetic energies are computed as an integral over the whole domain~$\Omega$:\\
\begin{equation}
P=\frac12\sum\limits^2_{i=1}\frac{H_{i} 
S_{i2}}{L^2 H}\iint\limits_{\Omega}(\psi_{i+1}-\psi_{i})^2\, dy dx \,, \quad H=\sum\limits^3_{i=1}H_{i} \,,
\label{eq:P}
\end{equation}
\begin{equation}
K=\frac12\sum\limits^3_{i=1}\frac{H_{i}}{A}\iint\limits_{\Omega}(\nabla\psi_{i})^2\, dy dx \,.
\label{eq:K}
\end{equation}
The layer-wise kinetic energy is given by\\
\begin{equation}
K_{i}=\frac12\frac{H_{i}}{A}\iint\limits_{\Omega}(\nabla\psi_{i})^2\, dy dx \,, \quad  i=1,2,3\,.
\label{eq:K_layer}
\end{equation}
Note that only $P$ and $K$ are included in the optimization procedure, i,e,
it can only guarantee that these quantities are within the reference energy band. It does not guarantee
that the layer-wise energy is within the band of the corresponding layer (and it is not; not shown). The layer-wise energy can be included in the optimization, but it is not covered in our study.
}

As seen in figure~\ref{fig:energy}, over the first approximately 38 years the reference solution $\mathbf{q}$ 
is in transit showing significant variations of kinetic $K(\mathbf{q})$ and potential, $P(\mathbf{q})$, energies; the low-resolution solution $\mathbf{q}^c$ (computed on the coarse grid) is equilibrated much faster though. The striking difference between these solutions is that 
$K(\mathbf{q}^c)$ and $P(\mathbf{q}^c)$ are significantly lower than $K(\mathbf{q})$ and $P(\mathbf{q})$. More importantly, 
$K(\mathbf{q}^c)$ and $P(\mathbf{q}^c)$ experience a sudden fall compared with $K(\mathbf{q})$ and $P(\mathbf{q})$, {\it when started from a reference initial condition} after the spinup. Eventually, both $K(\mathbf{q}^c)$ and $P(\mathbf{q}^c)$ reach even lower levels, which are natural to the low-resolution solution. This is typical and explained by the significant energy dissipation in the coarse-grid model and its impaired intra- and inter-scale energy transfers (including both forward and backward energy cascades). In phase space, this energy drop is observed as a drift of the low-resolution trajectory (solution) from the reference phase space, because the trajectory is gradually slipping off from the reference energy manifold.

\section{The hybrid quasi-geostrophic model}
In order to get the hybrid QG model, we add the velocity corrector $\mathbf{A}$ and the compensating forcing $\mathbf{G}$ to the QG model~\eqref{eq:pve}:\\
\begin{equation}
d q^h_j+(\mathbf{u}_j dt+\mathbf{A}(\mathbf{u}_j,\mathbf{v}_j))\cdot\nabla q^h_j=(F+\mathbf{G}(q^h_j,q_j))dt,
\label{eq:pve_hybrid}
\end{equation}
where $\mathbf{G}(q^h_j,q_j):=\eta(\mathbf{M}(q^h_j,q_j)-q^h_j)$, and the superscript $h$ denotes the hybrid solution. 
The system of elliptic equations~\eqref{eq:pv}, 
the mass conservation constraint~\eqref{eq:masscon}, and the boundary condition~\eqref{eq:bc} remain unchanged.

\ansA{
The resolution of the hybrid model is the same as for the QG model~\eqref{eq:pve}, i.e. 30~km.
It is in no way enough to resolve even the first Rossby radius ($Rd_1=40\, {\rm km}$) not to mention the second one ($Rd_2=23\, {\rm km}$). Thus, baroclinic instability is not explicitly resolved in the hybrid model. 
It makes this case harder compared to the one when the baroclinic instability is resolved,
as the hybrid model needs to both dissipate and inject energy where required.
}

We perform the multi-scale decomposition of the reference data $\{\mathbf{q}_i,\mathbf{v}_i\}_{i\in[1,N]}$. For the purpose of this study, it is enough to decompose the reference data into two fields: the first one includes all
scales within the range $s_1\in[30,300)\, {\rm km}$, and the second one includes all scales within the range $s_2\in[300,3840]\, {\rm km}$. It is done so, because the coarse-grid properly resolves scales larger
than $300\, {\rm km}$.
\ansA{All scales smaller than $300\, {\rm km}$ are deemed unresolved, as per the scale resolution criterion (10 points per grid scale to resolve it).}
Note that this decomposition into two fields might be not enough for other GFD models, as an accurate resolution of certain scales does not guarantee the proper operation of intra- and inter-scale energy transfers, and therefore more scales should be taken into account.

\begin{figure}[h]
\hspace*{-0.25cm}
\begin{tabular}{lll}
\begin{minipage}{0.355\textwidth}$q_1$\end{minipage}  & 
\hspace*{-0.4cm}\begin{minipage}{0.355\textwidth}$q_2$\end{minipage}  & 
\hspace*{-0.4cm}\begin{minipage}{0.4\textwidth}$q_3$\end{minipage}\\
\multicolumn{3}{l}{\begin{minipage}{1.0\textwidth}\includegraphics[scale=0.29]{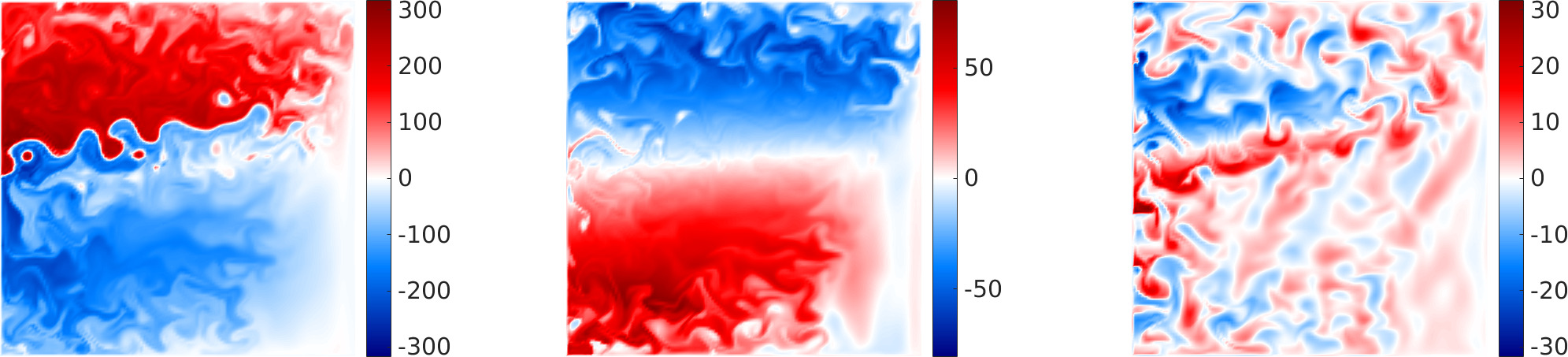}\end{minipage}} \\
\\[-0.125cm]
\begin{minipage}{0.355\textwidth}$s_1(q_1)$\end{minipage}  & 
\hspace*{-0.4cm}\begin{minipage}{0.355\textwidth}$s_1(q_2)$\end{minipage}  & 
\hspace*{-0.4cm}\begin{minipage}{0.34\textwidth}$s_1(q_3)$\end{minipage}\\
\multicolumn{3}{l}{\begin{minipage}{1.0\textwidth}\includegraphics[scale=0.29]{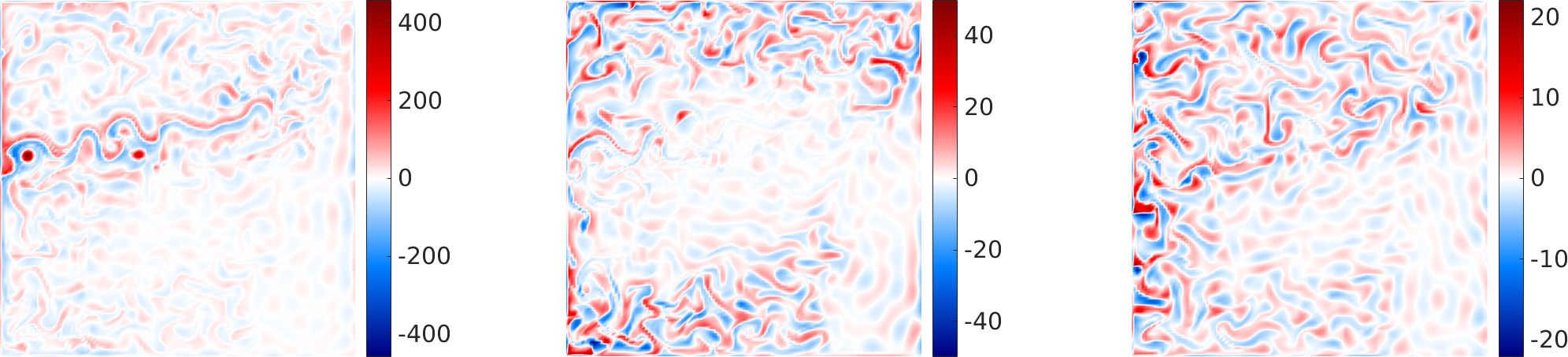}\end{minipage}} \\
\\[-0.125cm]
\begin{minipage}{0.355\textwidth}$s_2(q_1)$\end{minipage}  & 
\hspace*{-0.4cm}\begin{minipage}{0.355\textwidth}$s_2(q_2)$\end{minipage}  & 
\hspace*{-0.4cm}\begin{minipage}{0.34\textwidth}$s_2(q_3)$\end{minipage}\\
\multicolumn{3}{l}{\begin{minipage}{1.0\textwidth}\includegraphics[scale=0.29]{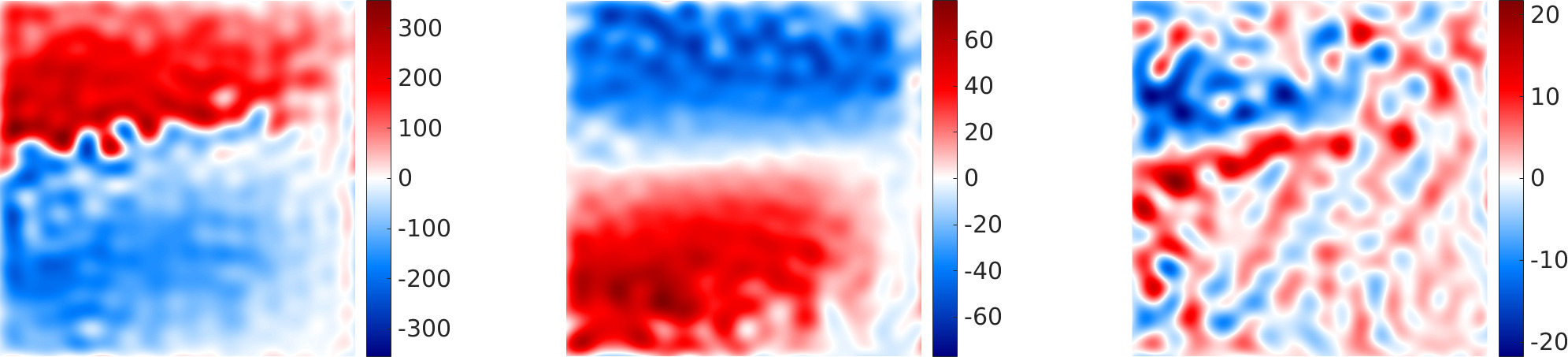}\end{minipage}}\\
\end{tabular}
\caption{Shown is a typical snapshot of the reference solution $\mathbf{q}$ on different levels,
and its multi-scale decomposition into scales $s_1\in[30,300)\, {\rm km}$ and $s_2\in[300,3840]\, {\rm km}$. }
\label{fig:multiscale_q}
\end{figure}

\begin{figure}[h]
\hspace*{-0.25cm}
\begin{tabular}{lll}
\begin{minipage}{0.355\textwidth}$q^c_1$\end{minipage}  & 
\hspace*{-0.4cm}\begin{minipage}{0.355\textwidth}$q^c_2$\end{minipage}  & 
\hspace*{-0.4cm}\begin{minipage}{0.34\textwidth}$q^c_3$\end{minipage}\\
\multicolumn{3}{l}{\begin{minipage}{1.0\textwidth}\includegraphics[scale=0.29]{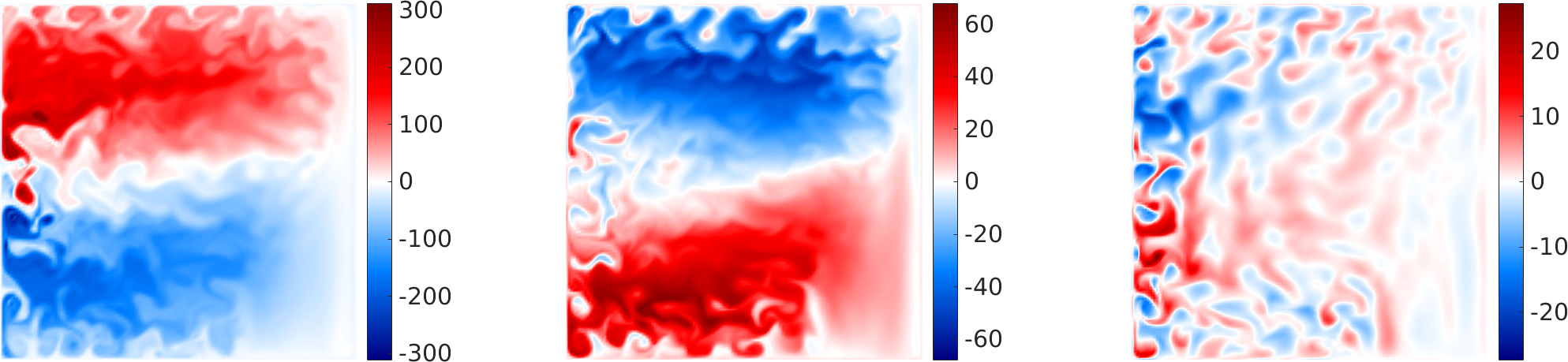}\end{minipage}} \\
\\[-0.125cm]
\begin{minipage}{0.355\textwidth}$s_1(q^c_1)$\end{minipage}  & 
\hspace*{-0.4cm}\begin{minipage}{0.355\textwidth}$s_1(q^c_2)$\end{minipage}  & 
\hspace*{-0.4cm}\begin{minipage}{0.34\textwidth}$s_1(q^c_3)$\end{minipage}\\
\multicolumn{3}{l}{\begin{minipage}{1.0\textwidth}\includegraphics[scale=0.29]{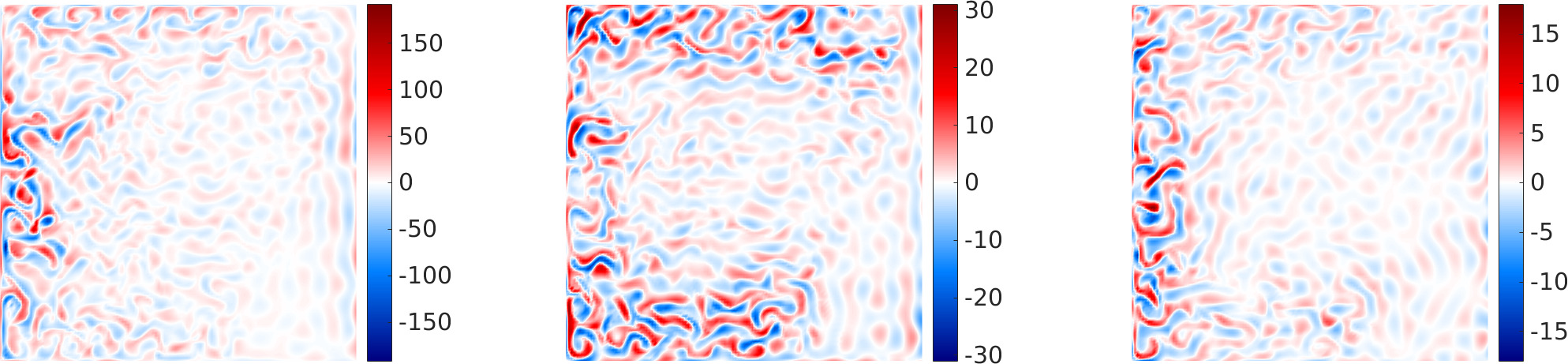}\end{minipage}} \\
\\[-0.125cm]
\begin{minipage}{0.355\textwidth}$s_2(q^c_1)$\end{minipage}  & 
\hspace*{-0.4cm}\begin{minipage}{0.355\textwidth}$s_2(q^c_2)$\end{minipage}  & 
\hspace*{-0.4cm}\begin{minipage}{0.34\textwidth}$s_2(q^c_3)$\end{minipage}\\
\multicolumn{3}{l}{\begin{minipage}{1.0\textwidth}\includegraphics[scale=0.29]{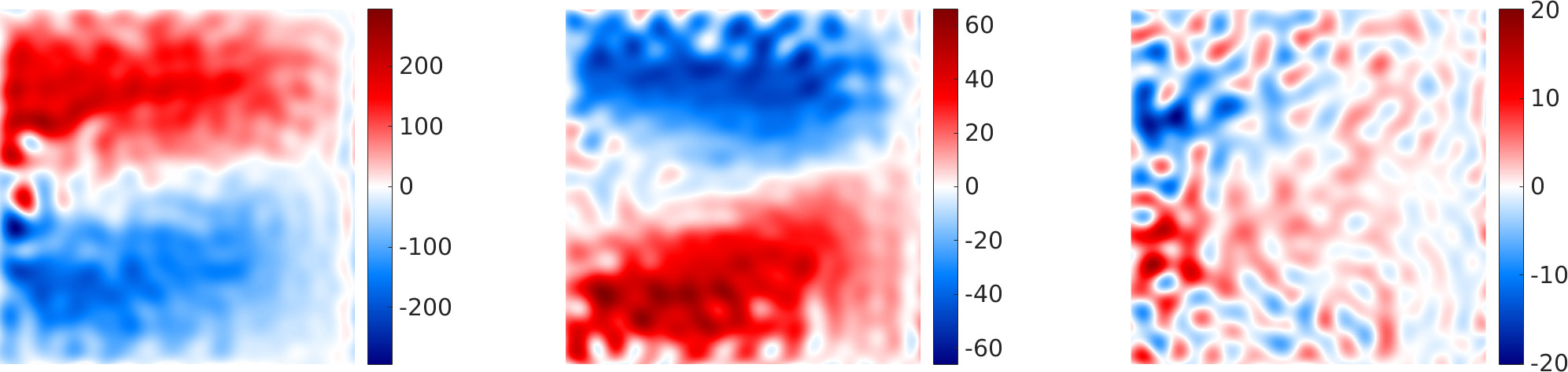}\end{minipage}}\\
\end{tabular}
\caption{The same as in figure~\ref{fig:multiscale_q} but for the low-resolution solution $\mathbf{q}^c$.}
\label{fig:multiscale_qc}
\end{figure}

We present the results of the multi-scale decomposition for both the reference (figure~\ref{fig:multiscale_q}) and low-resolution (figure~\ref{fig:multiscale_qc}) solutions to show how the lack of energy in the coarse-grid model influences the flow dynamics. 
As seen in figure~\ref{fig:multiscale_qc}, the coarse-grid model cannot reproduce the idealized Gulf Stream flow and the eddy-genesis is compromised, i.e. there are no eddies observed in the low-resolution solution. \\

The next step of the hybrid approach is to calculate the reference energy band for the kinetic and potential energies (from the 1-year reference record), which is $K(\mathbf{q})\in[76,90]$ and $P(\mathbf{q})\in[487,499]$, respectively. These boundaries are used in the optimization method to search for the scale amplitudes. 

Given the reference data, its multi-scale decomposition, and the reference energy band, we 
run the hybrid QG model~\eqref{eq:pve_hybrid} for a period of two years, and use only the first year of the reference data. 
For the purpose of this study, it is enough \ansA{to present the surface layer flow dynamics}, 
as it is much more energetic than the lower layers and full of both large- and small-scale flow features.

\section{Results}
We will consider different ideas on how to maintain the energy of the hybrid model within the reference energy band. 
Even though some failed, we think it is important for the reader to be aware of these failures to avoid them in the future.
The first one is to use the HP-like nudging~\cite{SB2021_J1} instead of the multi-scale forcing~\eqref{eq:m}:\\
\begin{equation}
\mathbf{G}^{\ast}(\mathbf{q}^h,\mathbf{q}):=\eta(\widehat{\mathbf{q}}-\mathbf{q}^h),\quad 
\widehat{\mathbf{q}}:=\frac{1}{M}\sum\nolimits_{i=\mathcal{U}_I}\mathcal{P}(\mathbf{q}_i)\Big|_{\mathcal{U}(\mathbf{q}^h(t,\cdot))},
\label{eq:nudge}
\end{equation}
which is equivalent to injecting or dissipating energy on all scales, as there is no multi-scale decomposition in use.

\begin{figure}[h!]
\hspace*{-0.5cm}
\begin{tabular}{ll}
\hspace*{0.2cm}(a)  & \hspace*{0.4cm}(b) \\
\begin{minipage}{0.5\textwidth}\includegraphics[scale=0.1325]{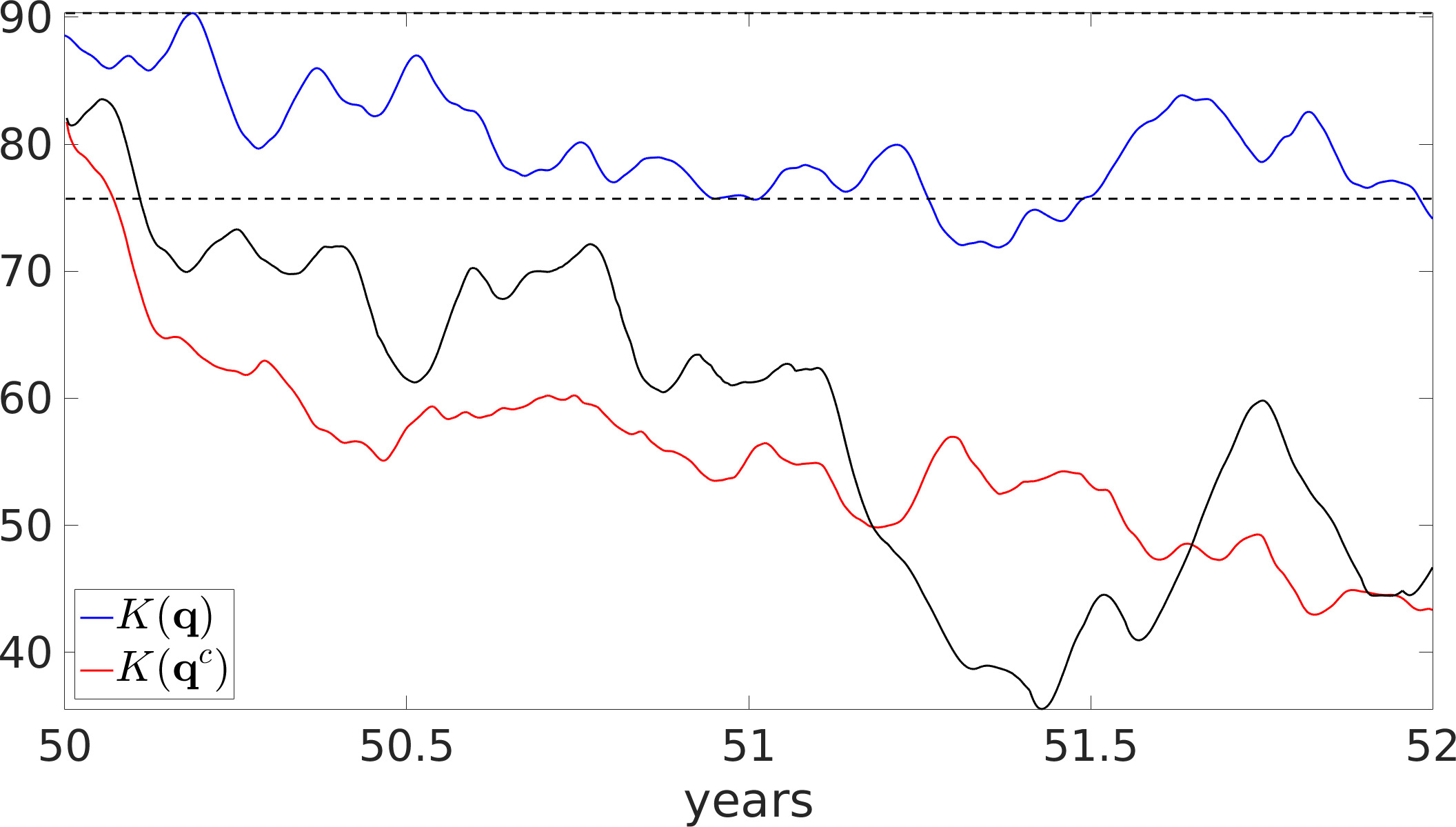}\end{minipage} &
\begin{minipage}{0.5\textwidth}\includegraphics[scale=0.13]{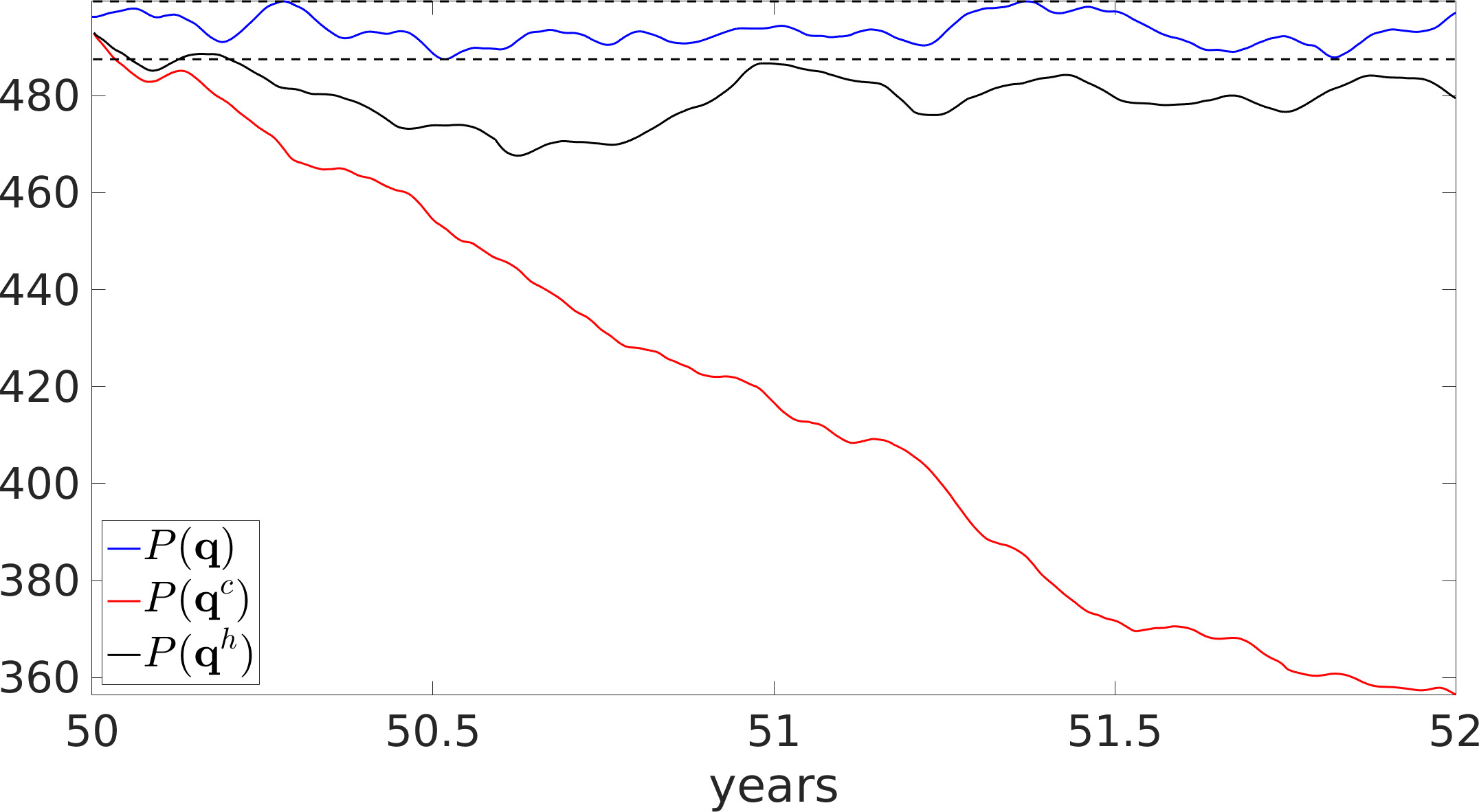}\end{minipage}\\
\end{tabular}

\hspace*{-0.75cm}
\begin{tabular}{llll}
& \begin{minipage}{0.355\textwidth}$q_1$\end{minipage}  & 
\hspace*{-0.4cm}\begin{minipage}{0.355\textwidth}$q^h_1$\end{minipage}  & 
\hspace*{-0.4cm}\begin{minipage}{0.34\textwidth}$q^c_1$\end{minipage}\\
\rotatebox{90}{t=0} & \multicolumn{3}{l}{\begin{minipage}{1.0\textwidth}\includegraphics[scale=0.29]{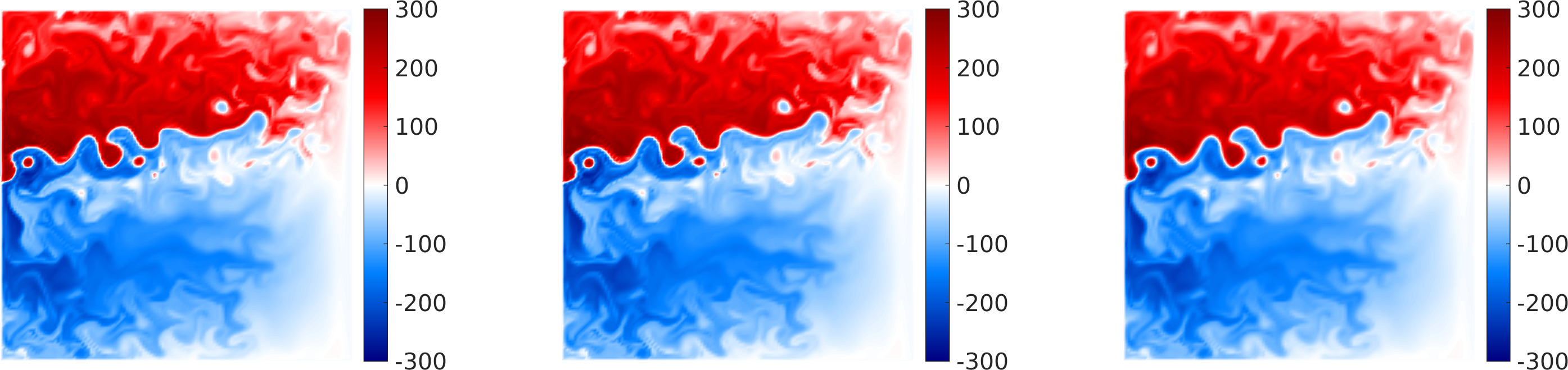}\end{minipage}} \\
& & &\\[-0.15cm]
\rotatebox{90}{\hspace*{-0.5cm}t=1 year} & \multicolumn{3}{l}{\begin{minipage}{1.0\textwidth}\includegraphics[scale=0.29]{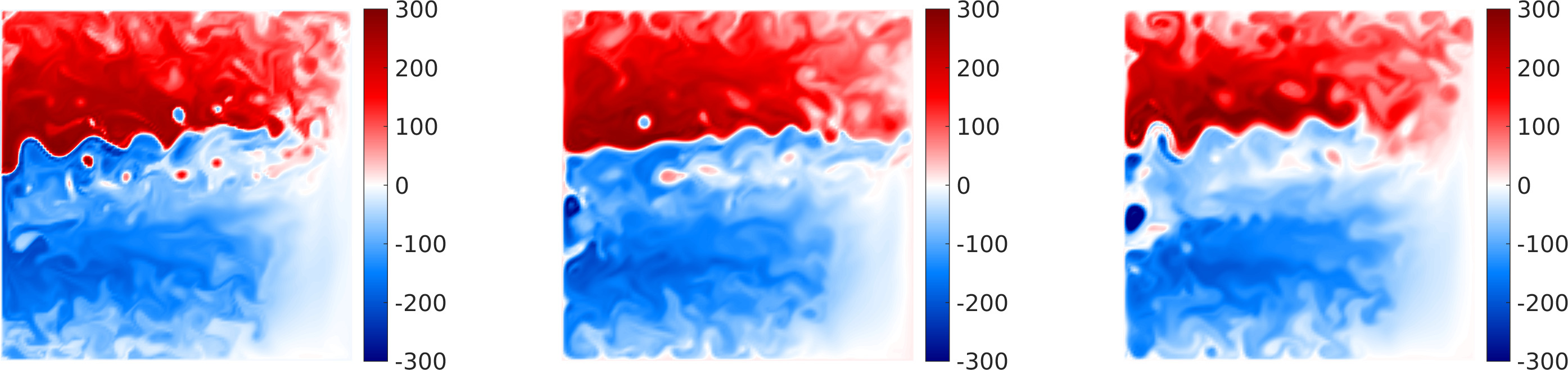}\end{minipage}} \\
& & &\\[-0.15cm]
\rotatebox{90}{\hspace*{-0.5cm}t=2 years} & \multicolumn{3}{l}{\begin{minipage}{1.0\textwidth}\includegraphics[scale=0.29]{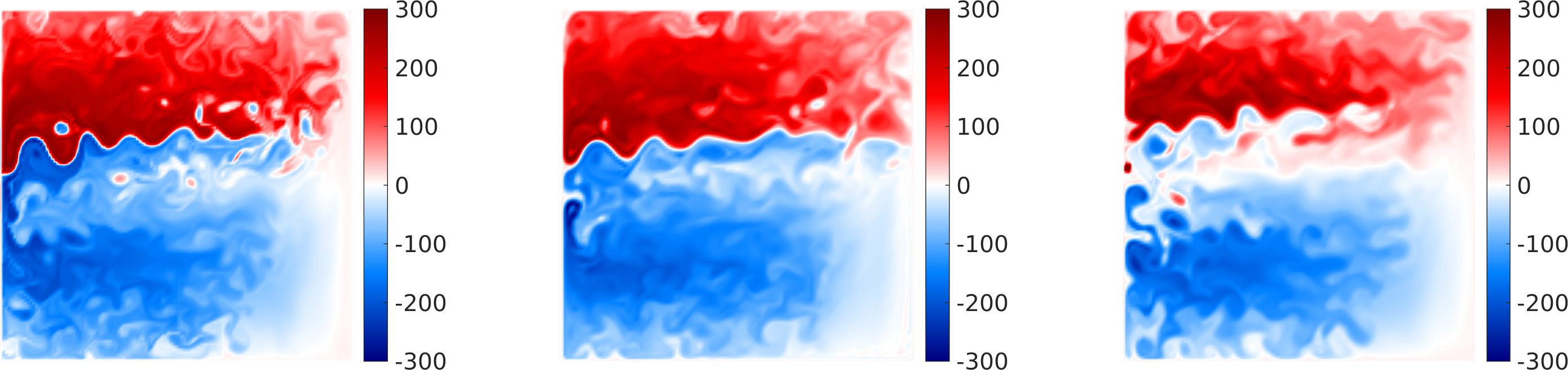}\end{minipage}} \\
\end{tabular}
\caption{Time series of the non-dimensional (a) kinetic and (b) potential energies of the reference $\mathbf{q}$ (blue),
low-resolution solution $\mathbf{q^c}$ (red), and hybrid solution $\mathbf{q^h}$ (black) for the case of using $\mathbf{G}^{\ast}$.
The kinetic and potential energies of the hybrid solution rapidly drop below the reference level (dashed line),
and even worse -- the former decreases lower than that of the low-resolution solution.
The subplots show snapshots of the reference solution $q_1$, hybrid solution $q^h_1$, and low-resolution solution $q^c_1$ 
at $t=\{0.0, 1.0, 2.0\}$ years; \ansA{the hybrid solution is computed for the HP-like nudging~\eqref{eq:nudge} which does not use the multi-scale decomposition.}  The lack of energy in the hybrid and low-resolution models lead to the loss of eddies and as a result to weakening the Gulf Stream flow.}
\label{fig:q_qh_qc_nudge}
\end{figure}

The detrimental effect of what happens to the kinetic and potential energies of the hybrid solution (figures~\ref{fig:q_qh_qc_nudge}a,b)
and what it does to the flow dynamics (figure~\ref{fig:q_qh_qc_nudge}) is clearly seen.
The energy drop in the hybrid and low-resolution models is accompanied by the degradation of eddy activity that in turn results in the inhibition of the Gulf Stream flow.
A similar situation (not shown) is observed for the case of fixed amplitudes $\lambda_s$ computed as the ratio between 
the time mean energies of the reference and low-resolution solutions on scales $s_1\in[30,300)\, {\rm km}$ and $s_2\in[300,3840]\, {\rm km}$.\\

\ansA{
In order to study how velocity correction and compensating forcing influences the flow dynamics and energy, 
it is important to consider the following deterministic ({\bf D}) and stochastic ({\bf S}) cases:\\
\begin{tabular}{llllll}
(i)$^*$ & ({\bf D}) & $\mathbf{A}\ne0$, $\mathbf{G}\ne0$ & $\mathbf{A}$ defined by~\eqref{eq:a} & criterion~\eqref{eq:opt2} & S=1\\  
(i) & ({\bf D}) & $\mathbf{A}=0$, $\mathbf{G}\ne0$ & $\mathbf{A}$ defined by~\eqref{eq:a} & criteria~\eqref{eq:opt1} \&~\eqref{eq:opt2} & S=2\\
(ii) & ({\bf D}) & $\mathbf{A}\ne0$, $\mathbf{G}=0$ & $\mathbf{A}$ defined by~\eqref{eq:a} & criterion~\eqref{eq:opt2} & S=2\\
(iii) & ({\bf D}) & $\mathbf{A}\ne0$, $\mathbf{G}\ne0$ & $\mathbf{A}$ defined by~\eqref{eq:a} & criterion~\eqref{eq:opt2} & S=2\\
(iv) & ({\bf S}) & $\mathbf{A}\ne0$, $\mathbf{G}\ne0$ & $\mathbf{A}$ defined by~\eqref{eq:dw} & criterion~\eqref{eq:opt2} & S=2
\end{tabular}
}

\ansA{
{\bf Deterministic case i$^*$} ($\mathbf{A}\ne0$ and $\mathbf{G}\ne0$; criterion~\eqref{eq:opt2};  $S=1$). In order to check whether the proposed multi-scale decomposition is really necessary, we consider the case with no multi-scale decomposition
and use criterion~\eqref{eq:opt2}. This criterion works better than criterion~\eqref{eq:opt1}, as verified in other cases below. We report the results for the case of no multi-scale decomposition (i.e. for $S=1$) in figure~\ref{fig:mean_energy0}. As seen in figures~\ref{fig:mean_energy0}a,b\,, 
the kinetic and potential energies of the hybrid solution (black line) experience significant unphysical fluctuations over the whole period of two years. The reason for these fluctuations is a constant competition between the hybrid model (which tries to 
leave the reference energy manifold by dissipating the excess of the reference total energy it cannot handle) and 
the optimization method (which constantly pushes the model back onto the reference energy manifold).
The result of this competition is a flow which is significantly corrupted by numerical instabilities appearing on the jet (figures~\ref{fig:mean_energy0}c,d) which can eventually lead to a numerical blow-up.
These instabilities come from the fact that the low-resolution QG model cannot cope with the whole chunk of the reference energy unless
this energy is decomposed into scales. This happens because both intra- and inter-scale energy transfers 
in the low-resolution model (including both forward and backward energy cascades) are impaired, and therefore the model cannot properly
redistribute the excess of the reference energy thus leading to numerical instabilities. 
This experiment (as well as the following ones) shows that the multi-scale decomposition is necessary to properly manage the reference energy in the hybrid model.\\
\begin{figure}[h]
\hspace*{-0.25cm}
\begin{tabular}{ll}
\hspace*{0.2cm}(a)  & \hspace*{0.2cm}(b) \\
\begin{minipage}{0.5\textwidth}\includegraphics[scale=0.13]{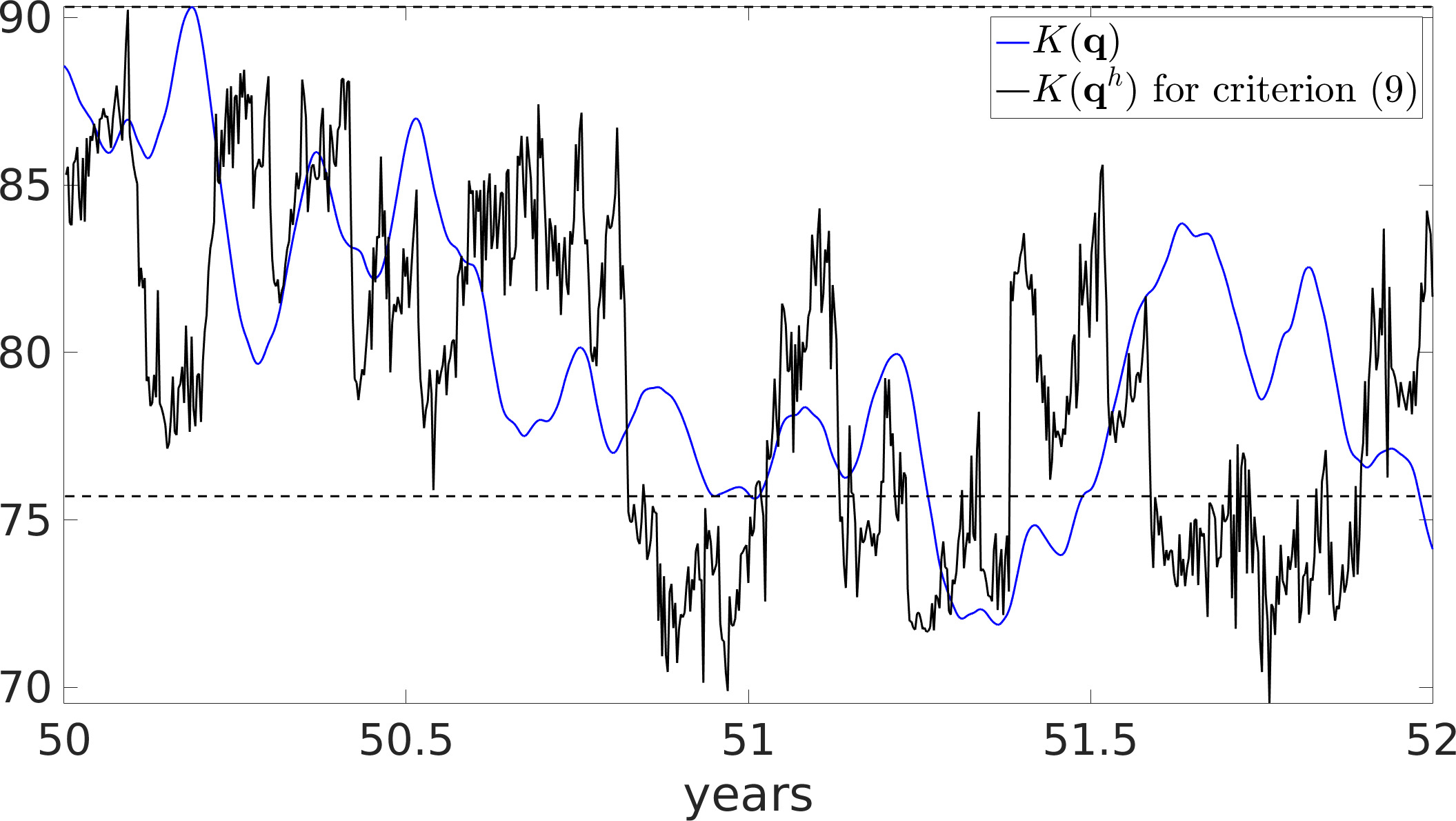}\end{minipage} &
\begin{minipage}{0.5\textwidth}\includegraphics[scale=0.13]{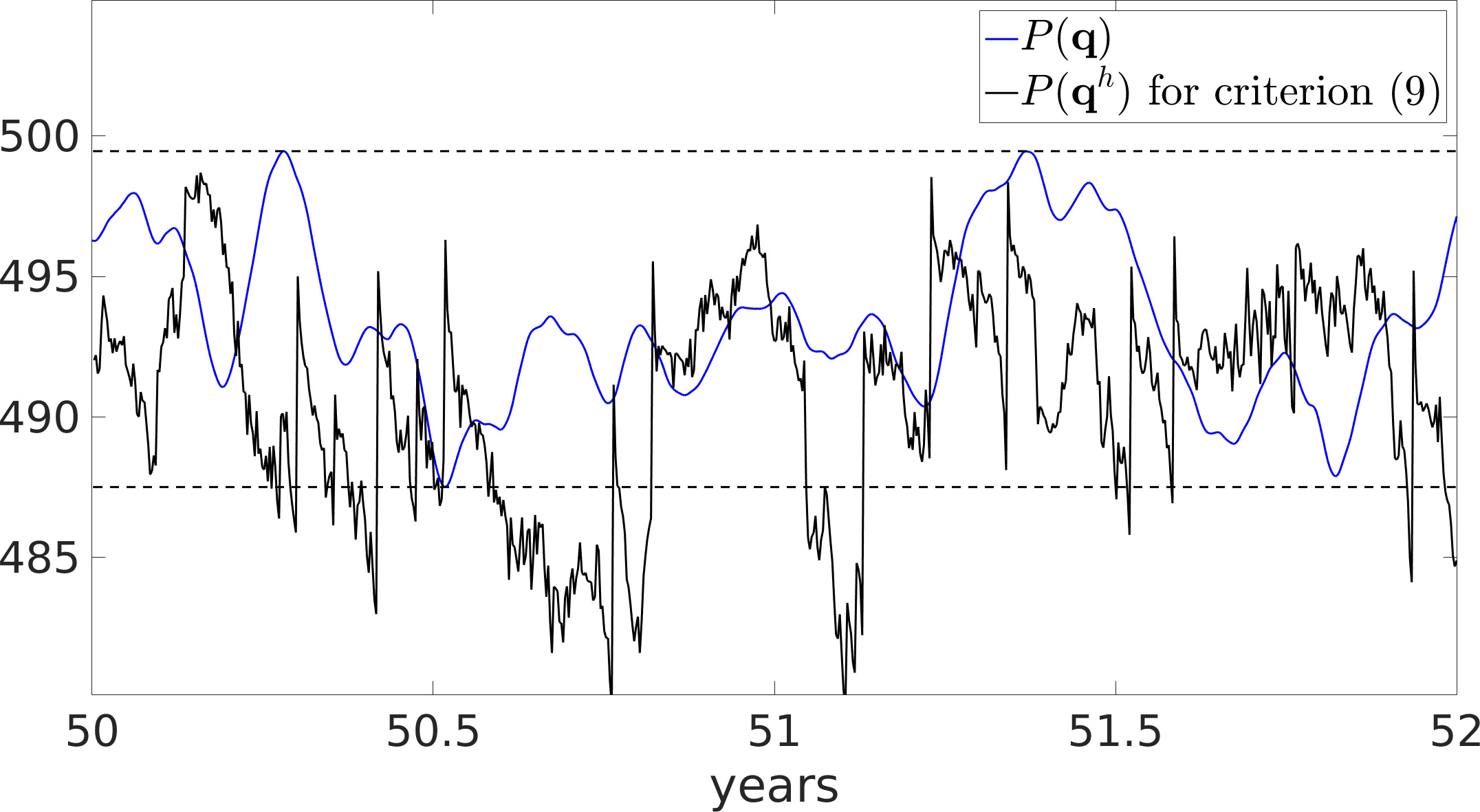}\end{minipage}
\end{tabular}\\[0.25cm]

\hspace*{-0.76cm}
\begin{tabular}{llll}
& \hspace*{0.125cm}$q^h_1$  & \hspace*{0.125cm}$q^h_1$   &  \hspace*{0.125cm}$q^h_1$  \\
\rotatebox{90}{(c)} & 
\begin{minipage}{0.33\textwidth}\includegraphics[scale=0.29]{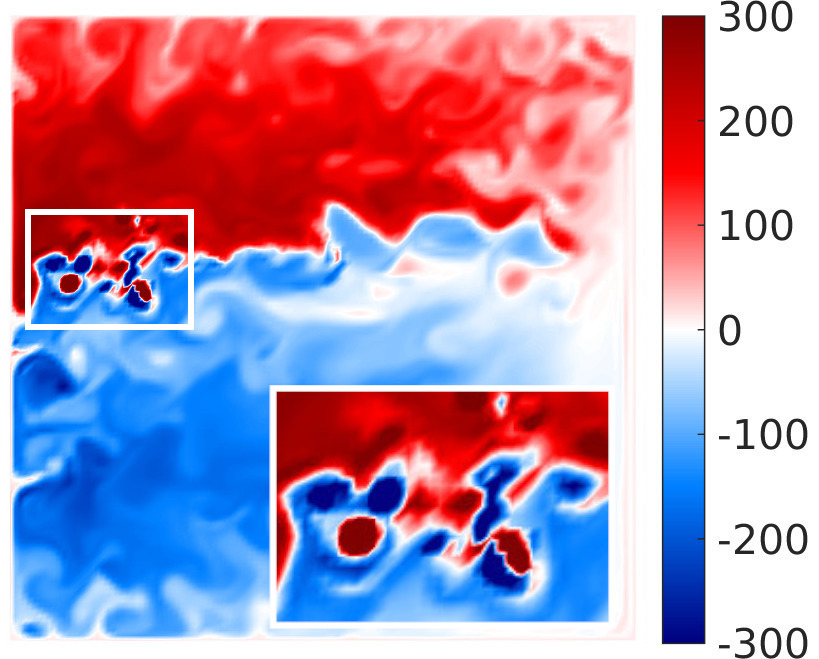}\end{minipage} & 
\begin{minipage}{0.33\textwidth}\includegraphics[scale=0.29]{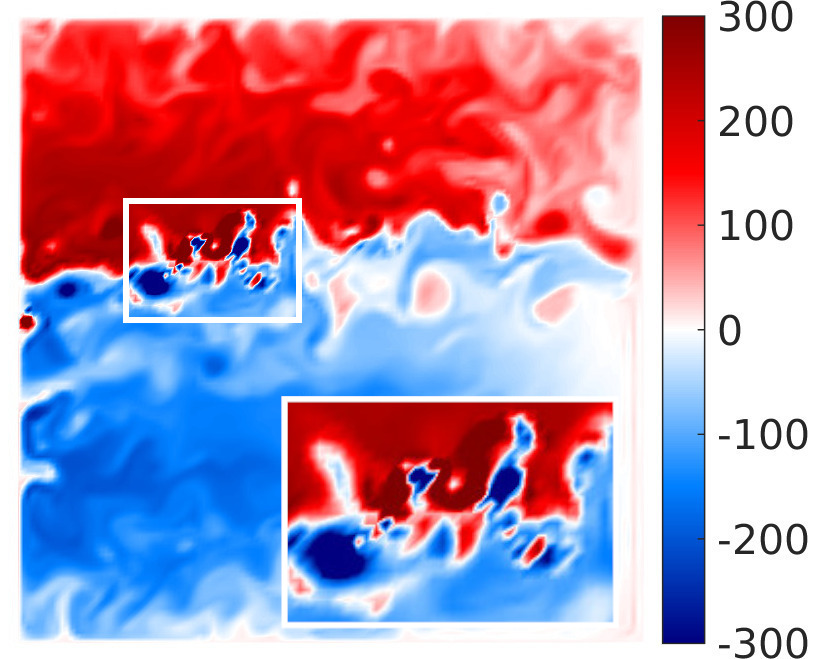}\end{minipage} &
\begin{minipage}{0.33\textwidth}\includegraphics[scale=0.29]{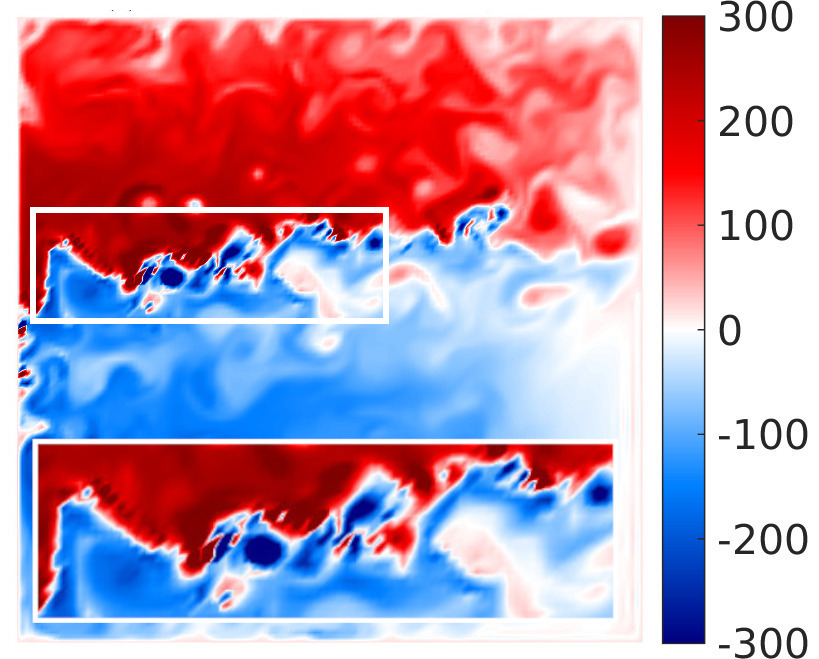}\end{minipage} \\
\end{tabular}\\[0.2cm]

\hspace*{-0.725cm}
\begin{tabular}{llll}
\rotatebox{90}{(d)} & 
\begin{minipage}{0.33\textwidth}\includegraphics[scale=0.29]{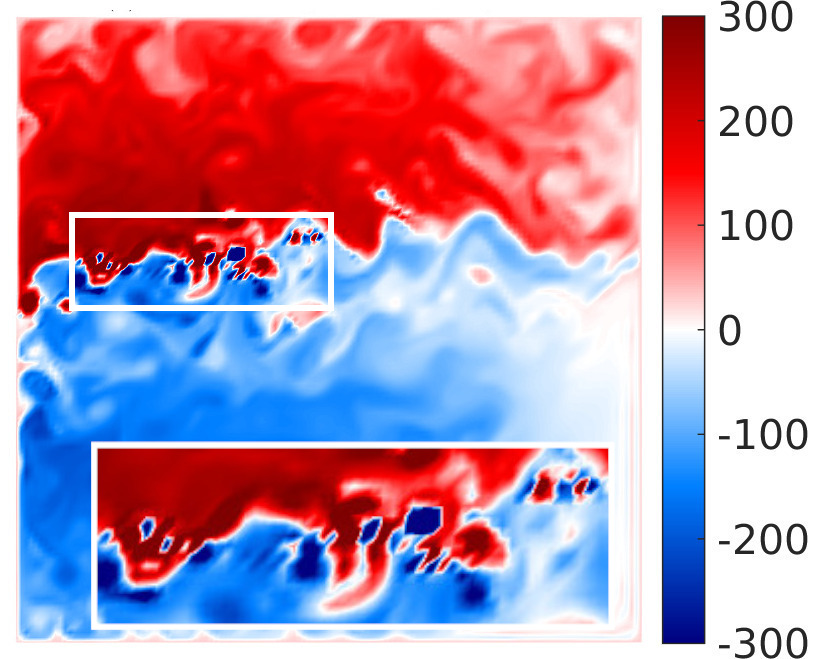}\end{minipage} & 
\begin{minipage}{0.33\textwidth}\includegraphics[scale=0.29]{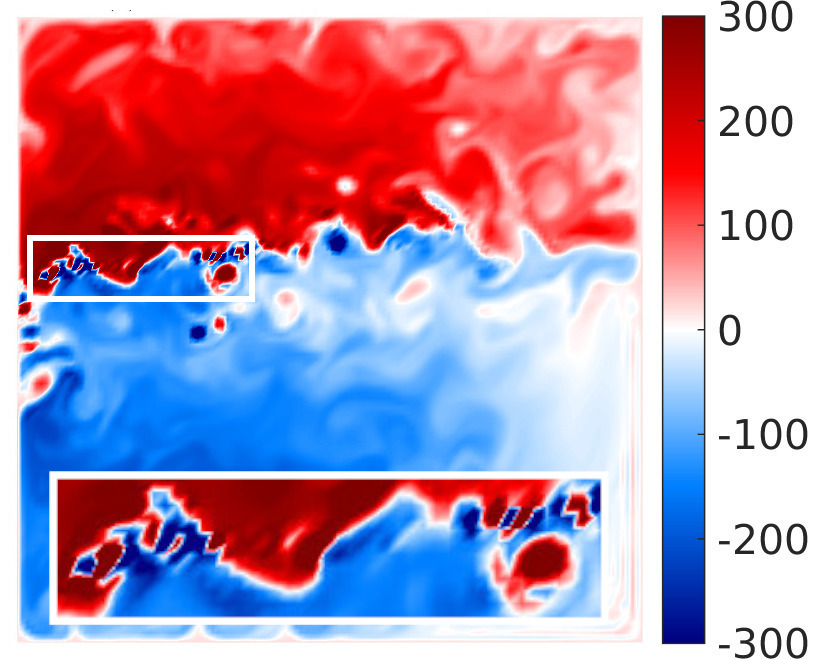}\end{minipage} &
\begin{minipage}{0.33\textwidth}\includegraphics[scale=0.29]{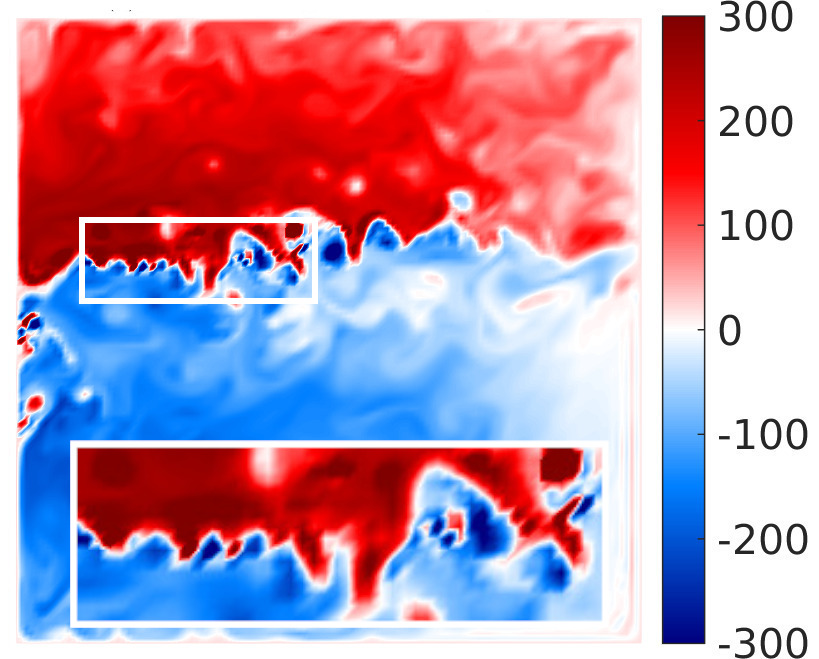}\end{minipage} \\
\end{tabular}

\caption{\ansA{Time series of the non-dimensional (a) kinetic and (b) potential energies of the reference $\mathbf{q}$ (blue line) and hybrid $\mathbf{q^h}$ (black line) solutions for criterion~\eqref{eq:opt2} and $S=1$ (no multi-scale decomposition used). Both the kinetic and potential energies of the hybrid solution experience significant unphysical fluctuations over the whole period of two years. These fluctuations leads to numerical instabilities (marked with a white rectangle in subplots (c) and (d)), which can result in a numerical blow-up.}
}
\label{fig:mean_energy0}
\end{figure}
}

\begin{figure}[h!]
\hspace*{-0.25cm}
\begin{tabular}{ll}
\hspace*{0.2cm}(a)  & \hspace*{0.2cm}(b) \\
\begin{minipage}{0.5\textwidth}\includegraphics[scale=0.13]{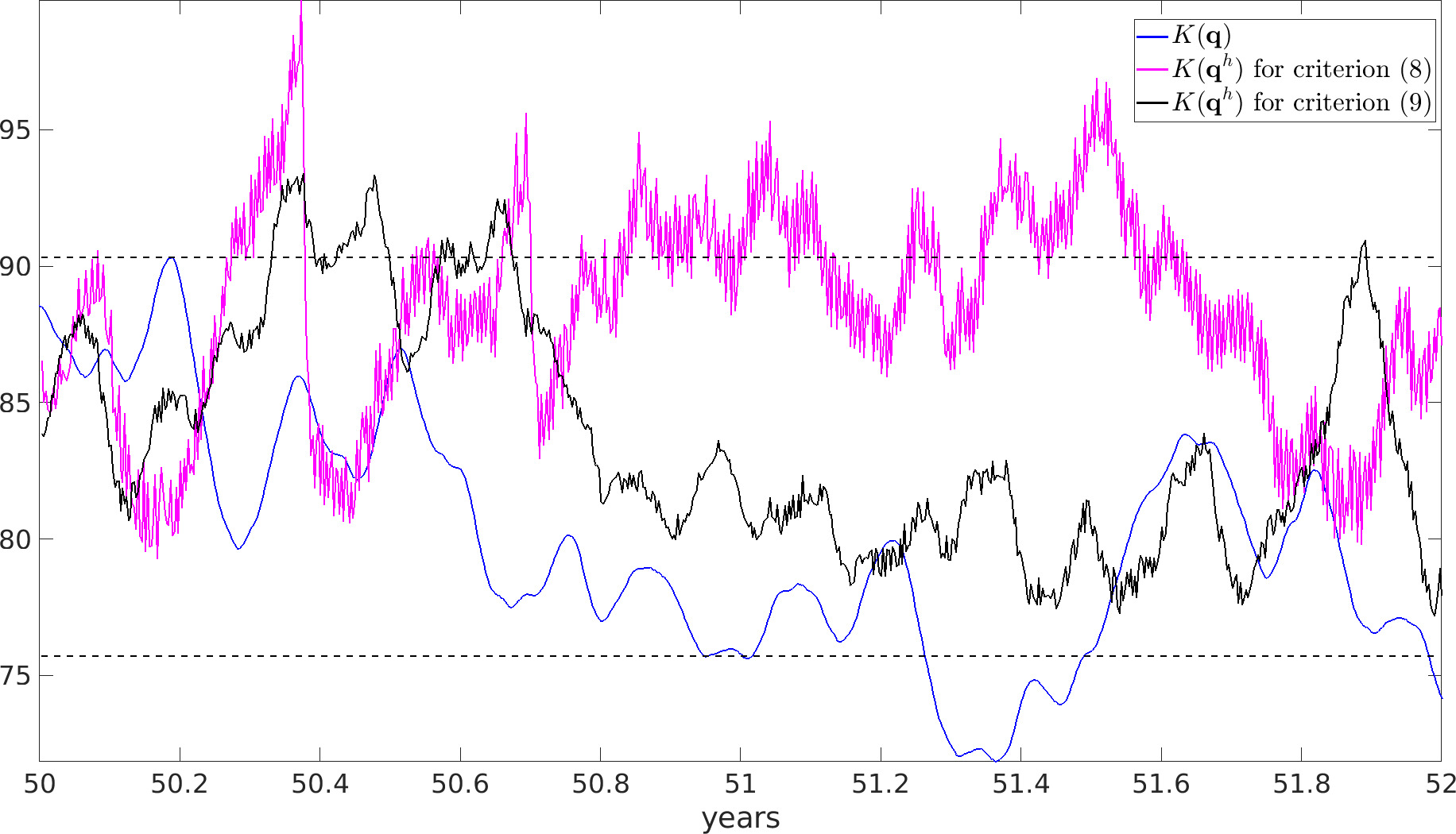}\end{minipage} &
\begin{minipage}{0.5\textwidth}\includegraphics[scale=0.13]{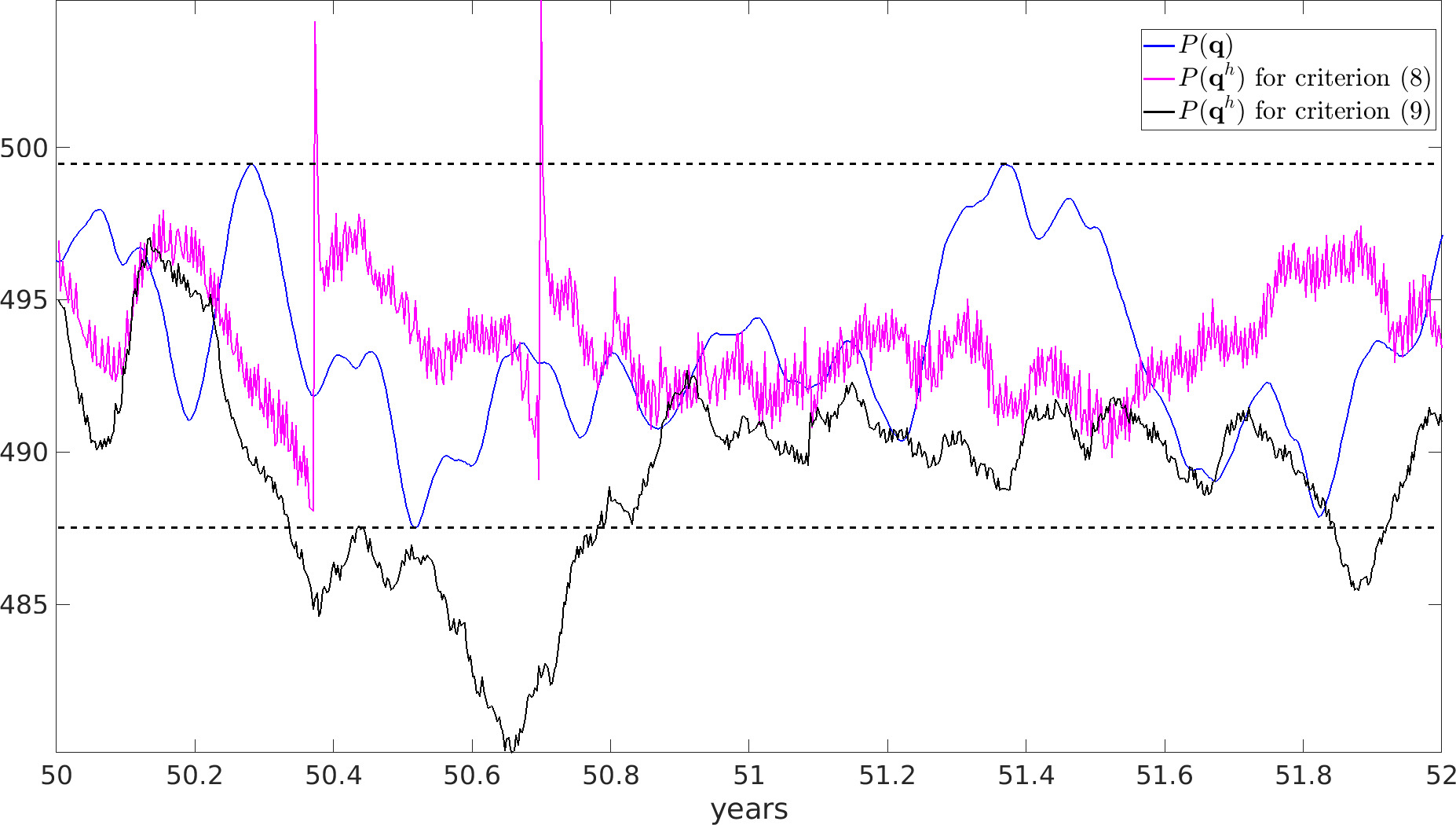}\end{minipage}
\end{tabular}\\[0.25cm]

\hspace*{-0.76cm}
\begin{tabular}{llll}
& \hspace*{0.125cm}$q^h_1$\hspace*{0.45cm} t=0.5 year & \hspace*{0.125cm}$q^h_1$\hspace*{0.45cm}  t=1.0 year  &  
\hspace*{0.125cm}$q^h_1$\hspace*{0.45cm}  t=2.0 years\\
\rotatebox{90}{(c)} & 
\begin{minipage}{0.33\textwidth}\includegraphics[scale=0.29]{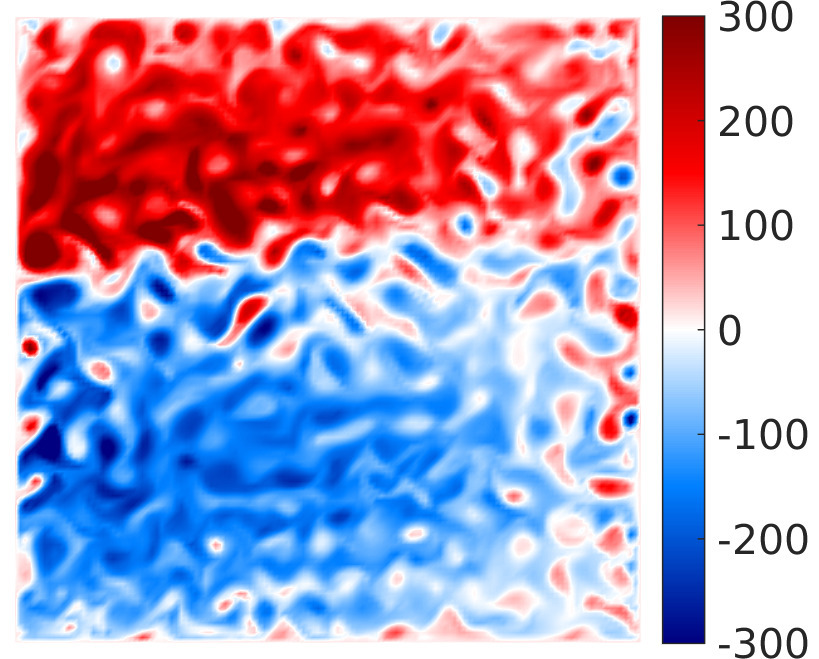}\end{minipage} & 
\begin{minipage}{0.33\textwidth}\includegraphics[scale=0.29]{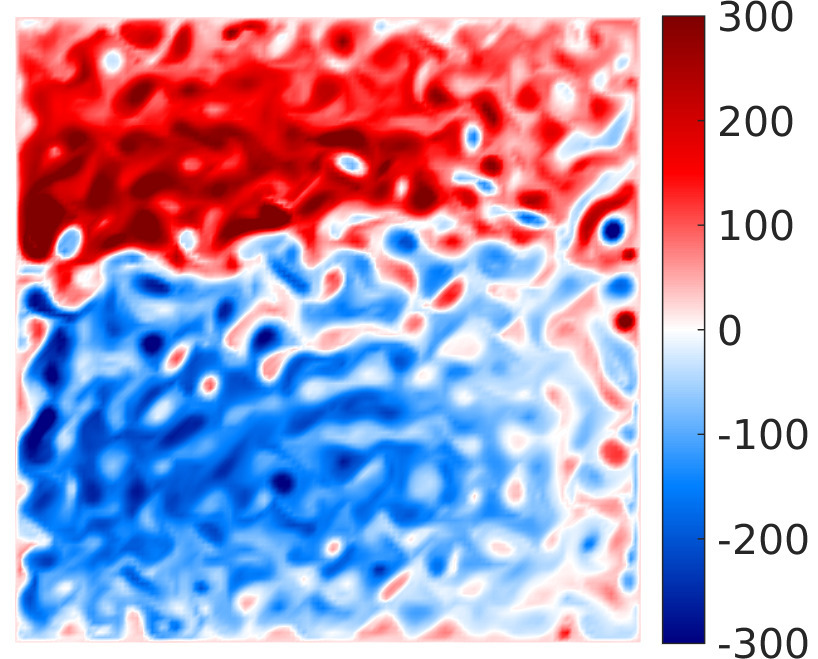}\end{minipage} &
\begin{minipage}{0.33\textwidth}\includegraphics[scale=0.29]{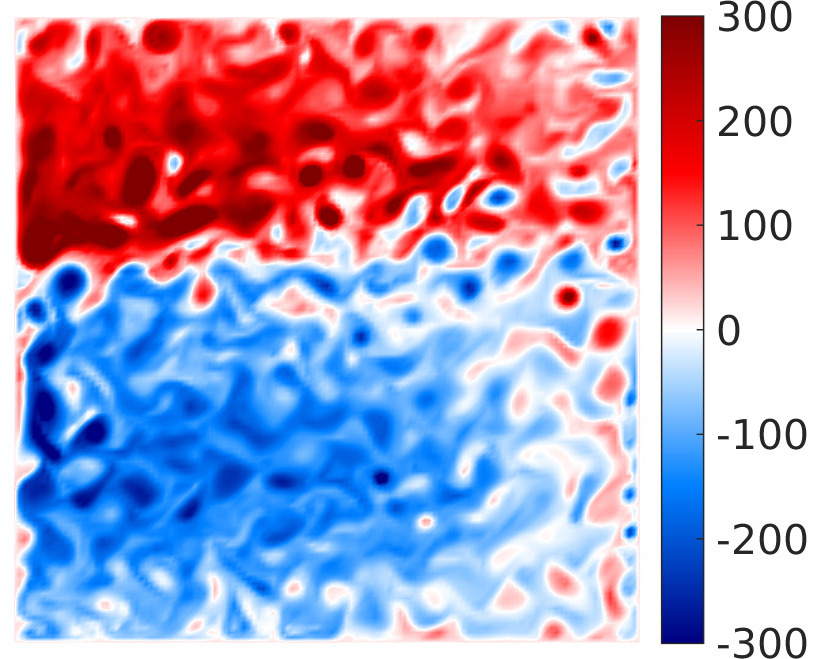}\end{minipage} \\
\end{tabular}\\[0.2cm]

\hspace*{-0.725cm}
\begin{tabular}{llll}
\rotatebox{90}{(d)} & 
\begin{minipage}{0.33\textwidth}\includegraphics[scale=0.29]{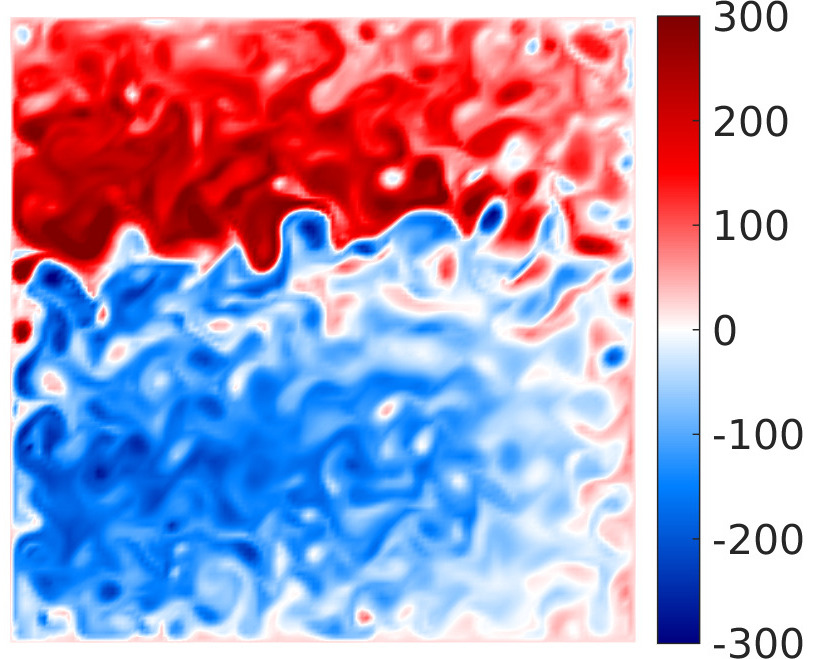}\end{minipage} & 
\begin{minipage}{0.33\textwidth}\includegraphics[scale=0.29]{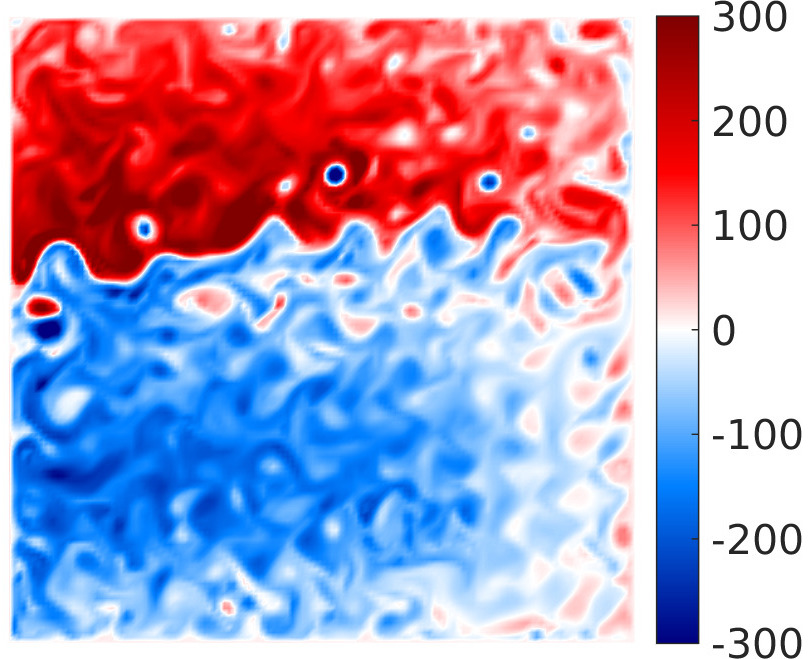}\end{minipage} &
\begin{minipage}{0.33\textwidth}\includegraphics[scale=0.29]{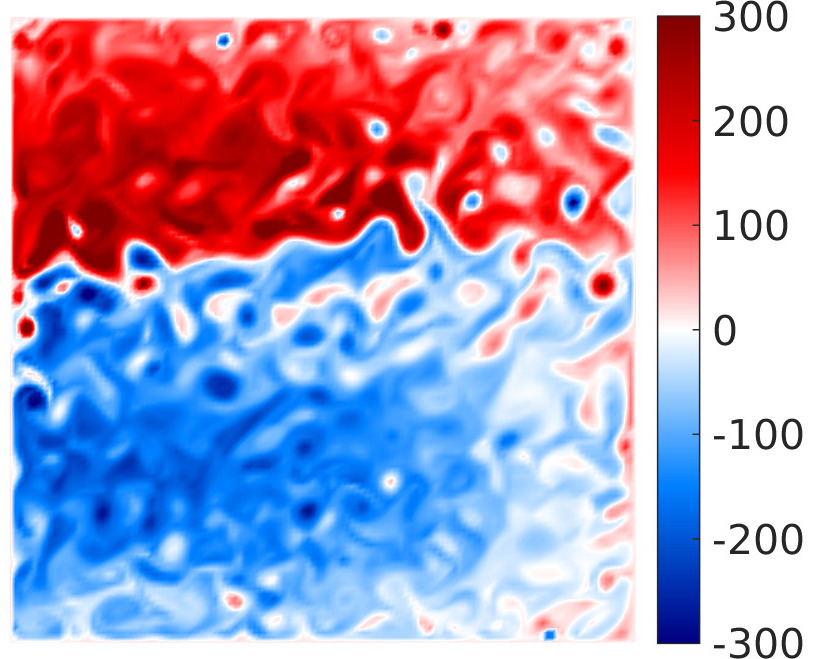}\end{minipage} \\
\end{tabular}

\caption{Time series of the non-dimensional (a) kinetic and (b) potential energies of the reference $\mathbf{q}$ (blue), and hybrid solution $\mathbf{q^h}$ for the first criterion~\eqref{eq:opt1} (magenta) and for the second criterion~\eqref{eq:opt2} (black). 
The use of the first criterion leads to appearing a high-frequency component in the kinetic and potential energies of the hybrid solution.
Moreover, the kinetic energy is above the reference level  (dashed line) thus resulting in the abundance of eddies and a less pronounced Gulf Stream flow (c). 
However, when the second criterion is in use, both the kinetic and potential energies of the hybrid solution are harmonized to the reference levels that results in significant improvements of the hybrid solution towards the reference flow (d).
}
\label{fig:mean_energy}
\end{figure}

{\bf Deterministic case i} ($\mathbf{A}=0$ and $\mathbf{G}\ne0$;  criteria~\eqref{eq:opt1} \&~\eqref{eq:opt2};  $S=2$).
To start with this case, we consider how the first optimization criterion~\eqref{eq:opt1} affects the energy and flow dynamics. As the results show 
(figures~\ref{fig:mean_energy}(a,b), magenta line), it brings the potential energy of the hybrid solution within the reference band, but the kinetic energy is still outside. 
Moreover, both energies experience high-frequency 
fluctuations due to constant (and significant) adjustments of the trajectory to satisfy the criterion. It means that the hybrid model is always trying to escape the reference energy manifold because the first criterion does not seem 
to properly correct intra- and inter-scale energy transfers towards the reference ones, which in turn leads to significant 
distortions of the flow dynamics along the jet (figure~\ref{fig:mean_energy}(c)) \ansA{and the eastern boundary.}
However, the flow dynamics improves dramatically if we use the second criterion~\eqref{eq:opt2}  (figure~\ref{fig:mean_energy}(d)).
Namely, the Gulf Stream flow becomes well-pronounced, teems with vortices, \ansA{and the activity of vortices along the eastern boundary is much weaker compared to the one in figure~\ref{fig:mean_energy}(c)}. 
Both the kinetic and potential energies of the hybrid solution are within the reference energy band,
although over a short time interval the potential energy experiences a burst out of the reference energy manifold. 
It happens because the optimization method cannot deliver the optimum after a fixed number of iterations used in this study, which is 10. \ansA{We expect an optimization over shorter time steps or
with more iterations should improve the situation, but we did not try it, as it is beyond the scope of this study.}\\

\begin{figure}[h!]
\hspace*{-0.25cm}
\begin{tabular}{ll}
\hspace*{0.4cm}(a)  & \hspace*{0.4cm}(b) \\
\begin{minipage}{0.5\textwidth}\includegraphics[scale=0.13]{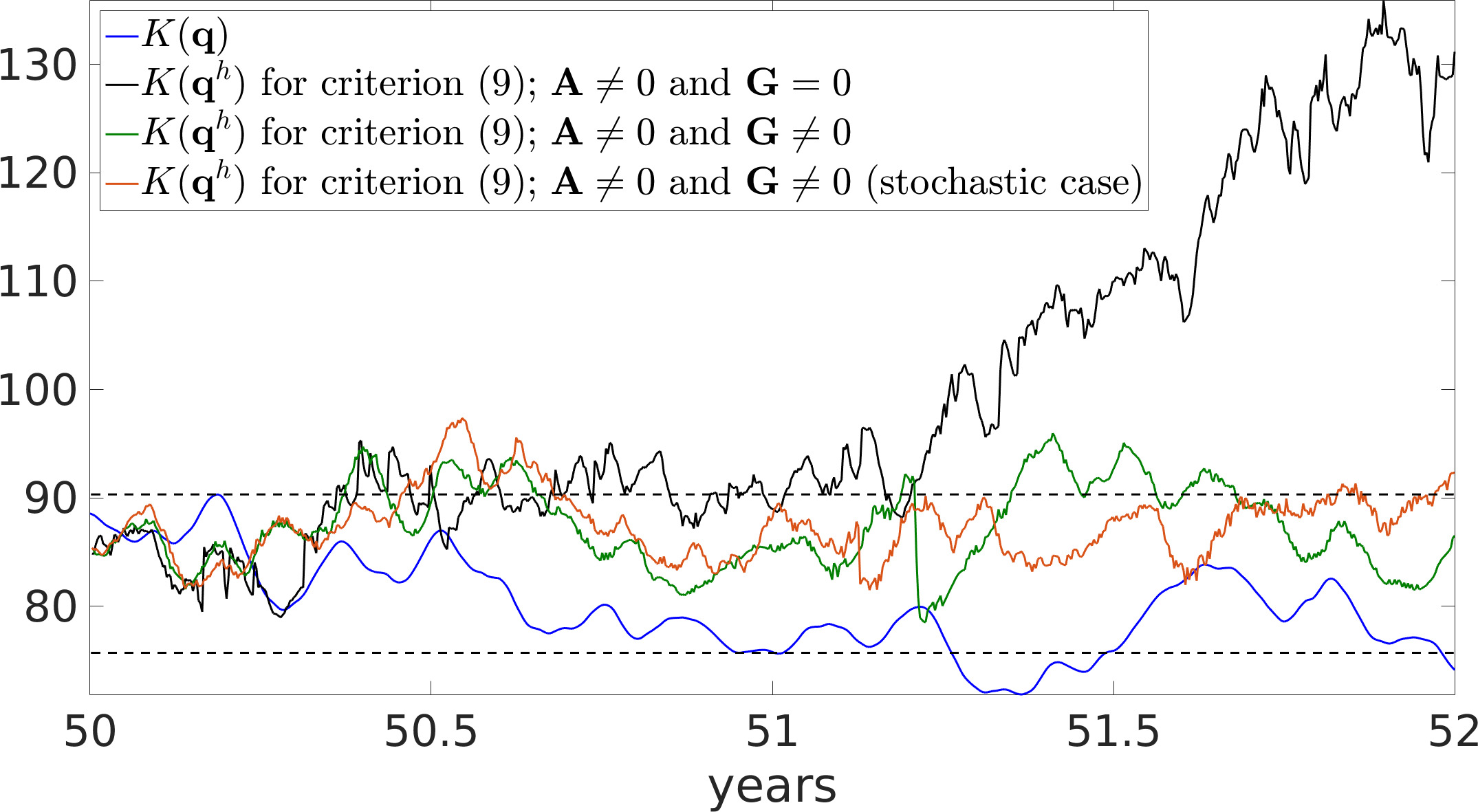}\end{minipage} &
\begin{minipage}{0.5\textwidth}\includegraphics[scale=0.13]{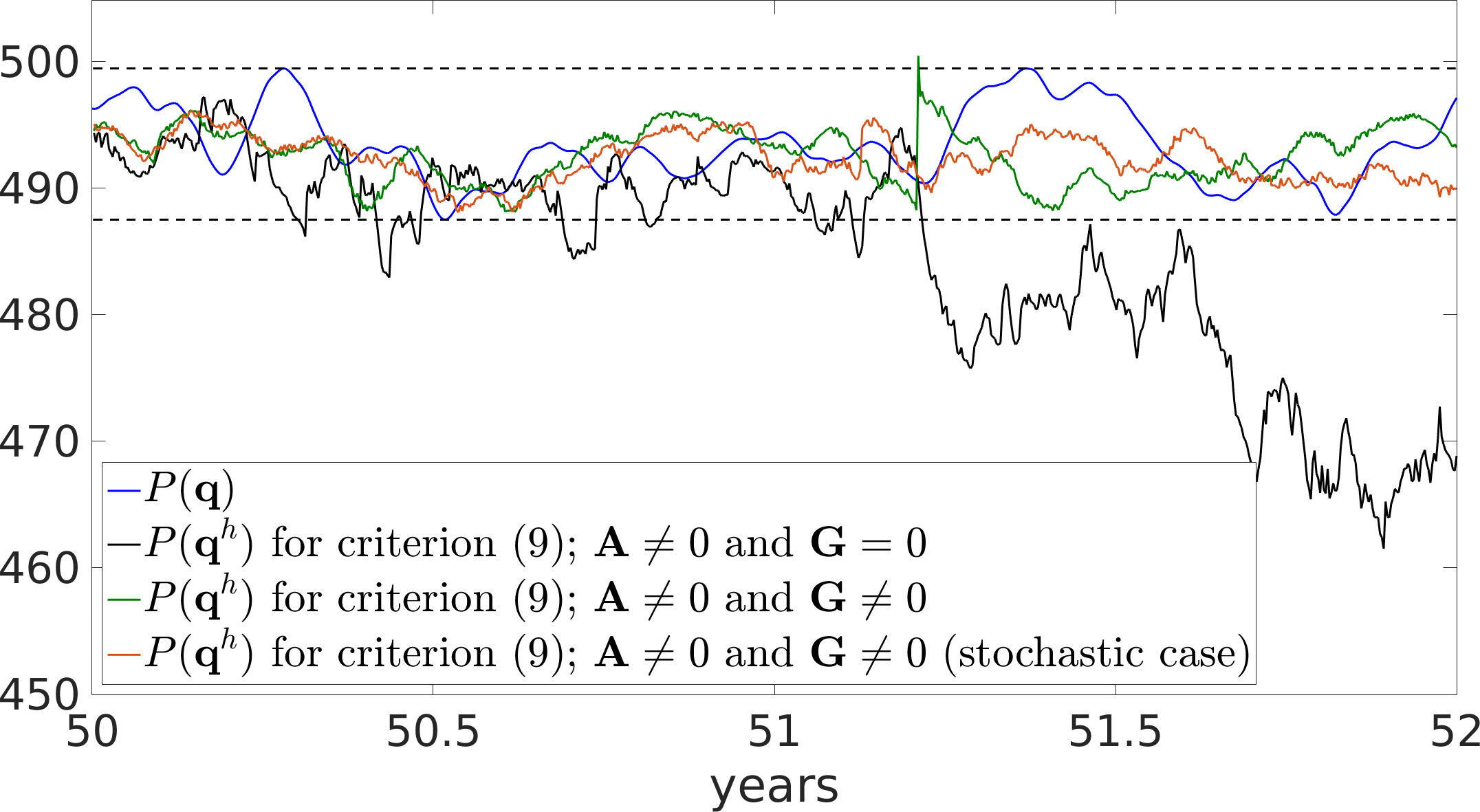}\end{minipage}
\end{tabular}\\[0.25cm]

\hspace*{-0.7cm}
\begin{tabular}{llll}
& \hspace*{0.125cm}$q^h_1$\hspace*{0.45cm} t=0.5 year & \hspace*{0.125cm}$q^h_1$\hspace*{0.45cm}  t=1.0 year  &  
\hspace*{0.125cm}$q^h_1$\hspace*{0.45cm}  t=2.0 years\\
\rotatebox{90}{(c)} & 
\begin{minipage}{0.33\textwidth}\includegraphics[scale=0.29]{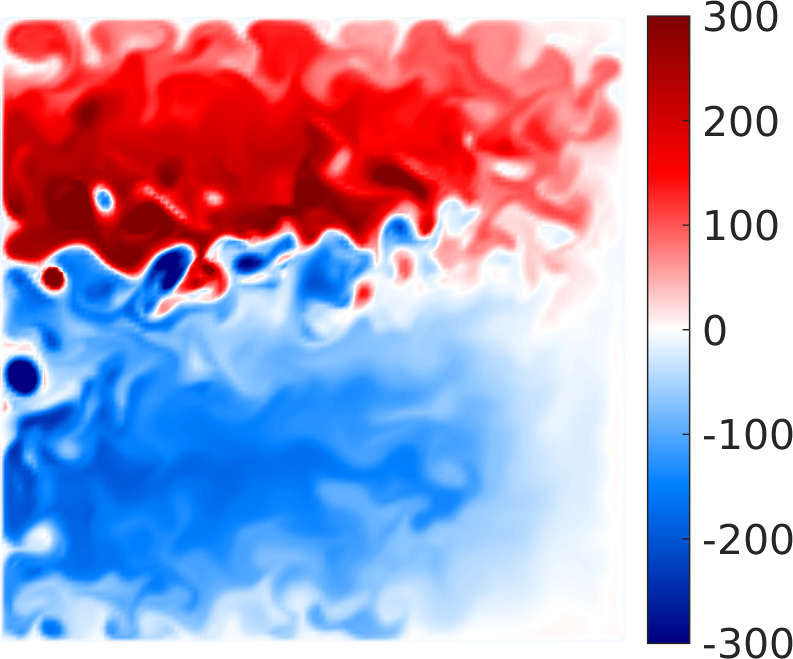}\end{minipage} & 
\begin{minipage}{0.33\textwidth}\includegraphics[scale=0.29]{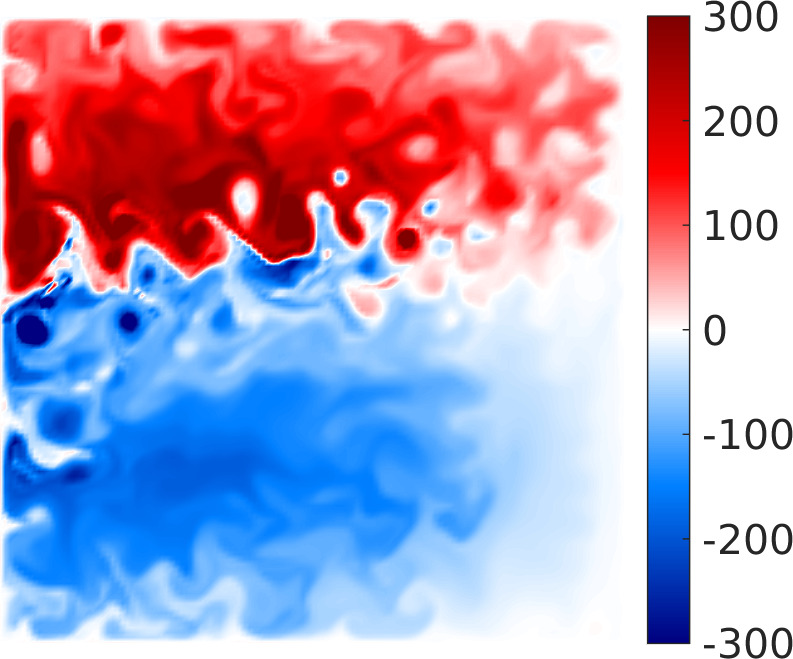}\end{minipage} &
\begin{minipage}{0.33\textwidth}\includegraphics[scale=0.29]{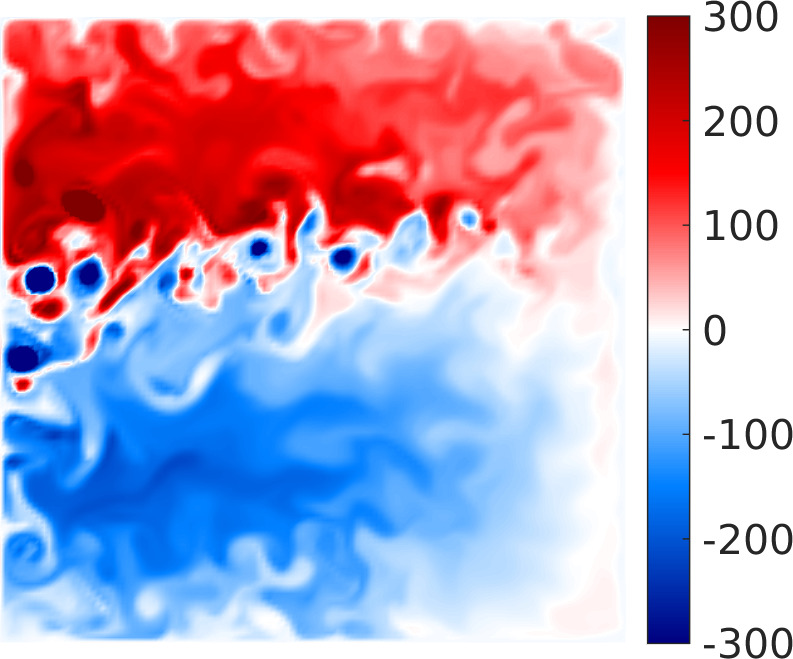}\end{minipage} \\
\end{tabular}\\[0.2cm]

\hspace*{-0.7cm}
\begin{tabular}{llll}
\rotatebox{90}{(d)} & 
\begin{minipage}{0.33\textwidth}\includegraphics[scale=0.29]{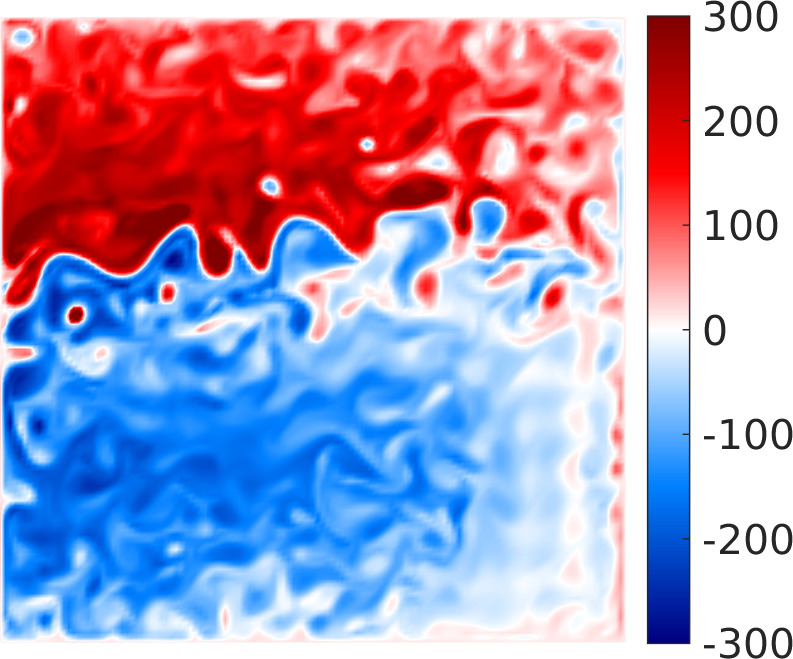}\end{minipage} & 
\begin{minipage}{0.33\textwidth}\includegraphics[scale=0.29]{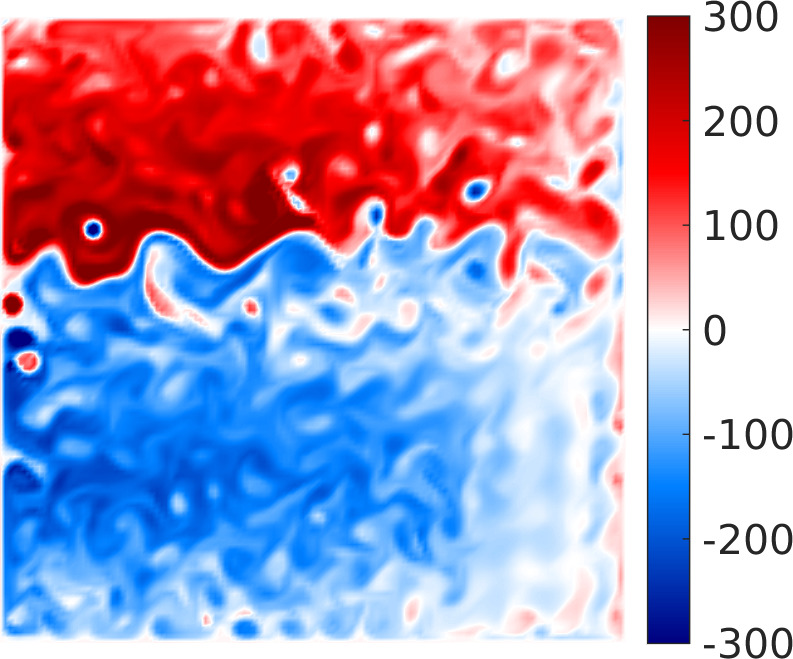}\end{minipage} &
\begin{minipage}{0.33\textwidth}\includegraphics[scale=0.29]{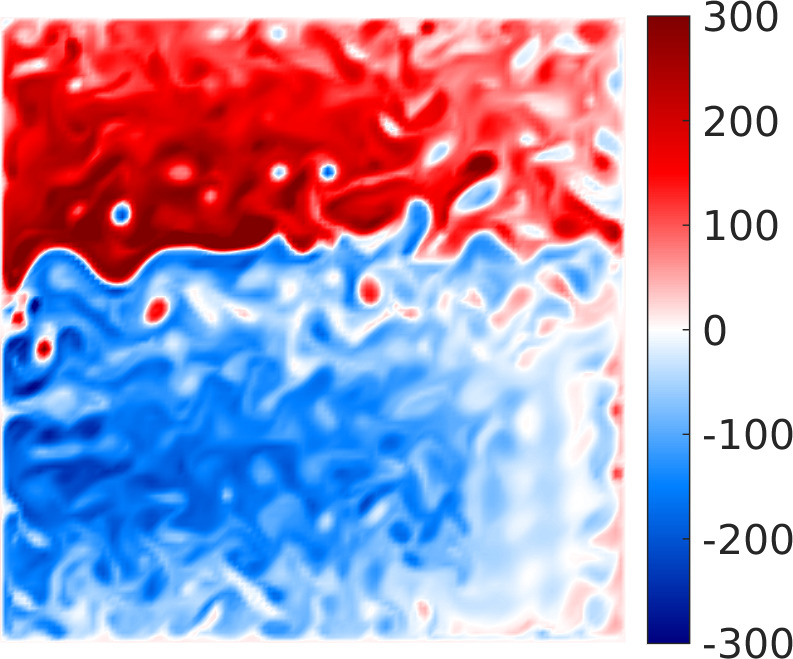}\end{minipage} \\
\end{tabular}\\[0.2cm]

\hspace*{-0.7cm}
\begin{tabular}{llll}
\rotatebox{90}{(e)} & 
\begin{minipage}{0.33\textwidth}\includegraphics[scale=0.29]{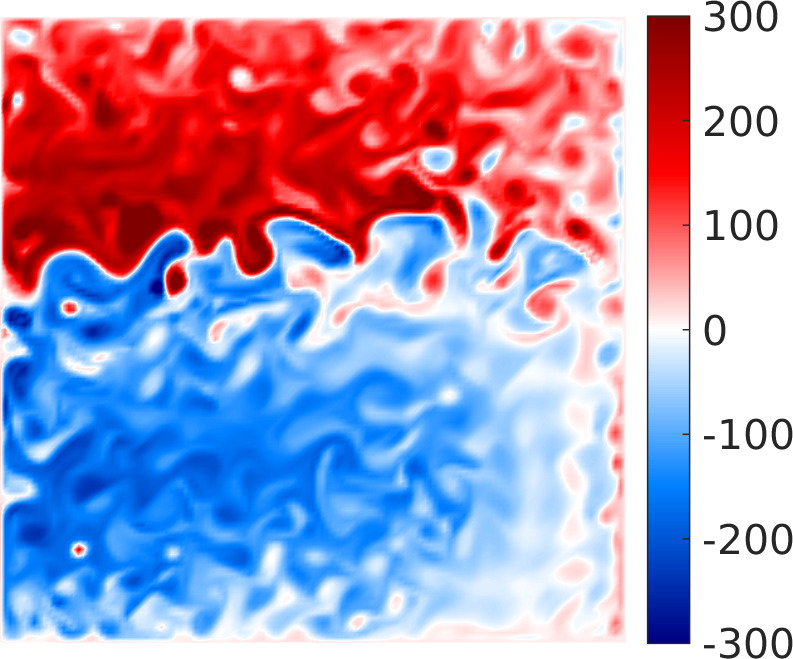}\end{minipage} & 
\begin{minipage}{0.33\textwidth}\includegraphics[scale=0.29]{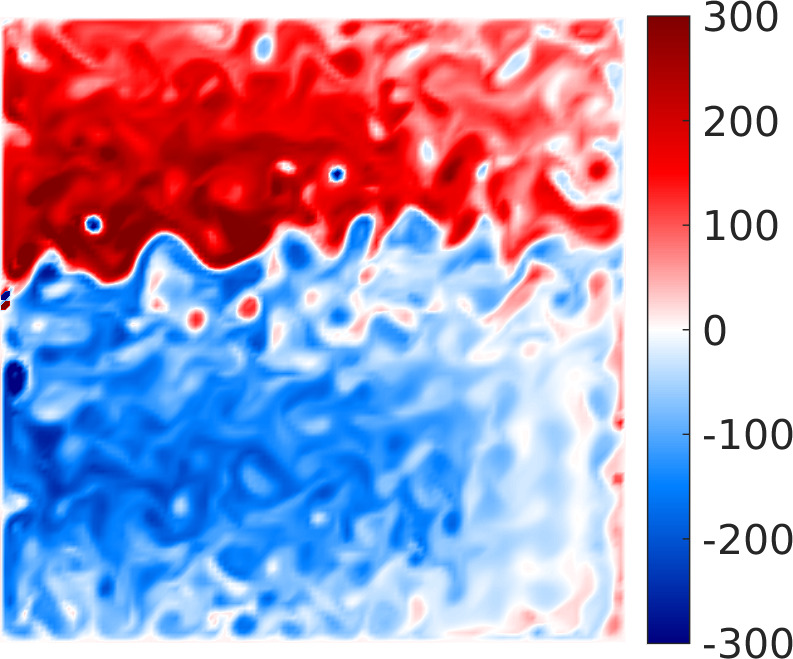}\end{minipage} &
\begin{minipage}{0.33\textwidth}\includegraphics[scale=0.29]{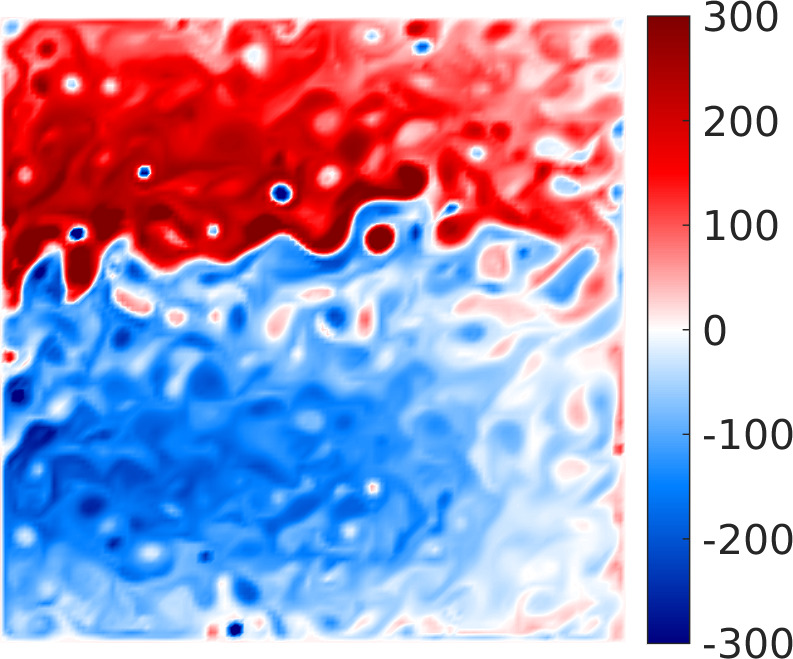}\end{minipage} \\
\end{tabular}

\caption{Time series of the non-dimensional (a) kinetic and (b) potential energies of the reference $\mathbf{q}$ (blue), and hybrid solution $\mathbf{q^h}$ for the second criterion~\eqref{eq:opt2}: case (ii) $\mathbf{A}\ne0$ and $\mathbf{G}=0$ (black);
(iii) $\mathbf{A}\ne0$ and $\mathbf{G}\ne0$ (green); and case (iv) stochastic version of case (iii) (orange). 
For case (ii), when the compensating forcing is disabled,
both kinetic and potential energy is well beyond the reference energy band which results in a significant corruption of the Gulf Stream flow along the jet (c). For case (iii), when the compensating forcing is enabled, the kinetic and potential
energies (green) are within the reference energy band, and the Gulf Stream flow dynamics is improved towards the reference one (d).
For case (iv), when the stochastic version of the velocity corrector $\mathbf{A}$ is used together with the compensating forcing $\mathbf{G}$, 
the kinetic and potential energies (orange) are even more accurate compared to all other cases, although the flow dynamics (e) does not manifest qualitative differences compared to case (iii). \ansA{Note that for both (iii) and (iv) the eddy activity
along the eastern boundary is even weaker than for case (ii), see figure~\ref{fig:mean_energy}(c), i.e. the solution of the hybrid model
further improves toward the reference one.}
}
\label{fig:mean_energy_2}
\end{figure}

{\bf Deterministic case ii} ($\mathbf{A}\ne0$ and $\mathbf{G}=0$; criterion~\eqref{eq:opt2}; $S=2$).
When the compensating forcing is disabled ($\mathbf{G}=0$), the optimization procedure
is unable to keep the energy of the hybrid model within the reference band
thus leading to significant bursts of energy out of the reference energy manifold (figures~\ref{fig:mean_energy_2}(a,b), black line)
that, in turn, leads to significant distortions to the jet itself (figures~\ref{fig:mean_energy_2}(c)).\\

{\bf Deterministic case iii} ($\mathbf{A}\ne0$ and $\mathbf{G}\ne0$; criterion~\eqref{eq:opt2}; $S=2$). 
The situation changes rapidly for the better if the compensating forcing is allowed. Namely, the energy of the hybrid model evolves in the reference energy band
(figures~\ref{fig:mean_energy_2}(a,b), green line) and the flow dynamics shows no sign of jet or vortices degradation (figure~\ref{fig:mean_energy_2}(d)). \ansA{Moreover, the eddy activity along the eastern boundary is even weaker than for case (ii), see figure~\ref{fig:mean_energy}(c), thus showing further improvement of the hybrid solution towards the reference one.
Besides, our ensemble simulations show that the hybrid model
is systematically more accurate (closer to the reference solution over the validation period) than the the low-resolution QG model (see stochastic ensemble simulations below).
We do not expect the reference and hybrid solutions to be perfectly alike, there will always be some differences and imperfections.

The energy of the deterministic hybrid solution might experience rapid fluctuations
like the one at $\sim$51.2 years (figures~\ref{fig:mean_energy_2}(a,b), green line).
We associate this rapid change of energy
with a lack of data in between different parts of the reference phase space, i.e. when the hybrid model is in what we call "void" (a region in the reference phase space which does not have enough reference data to be accurately represented), see figure~\ref{fig:ref_space3}. When the hybrid deterministic model is
in a void, and a correction to the reference energy band is needed, the model relies on the information (about the reference solution) which is available only on the boundary of the void. It can therefore lead to a rapid energy fluctuation if the void region is relatively large and the energy changes significantly across it.
This situation is resolved when using the stochastic hybrid model (figures~\ref{fig:mean_energy_2}(a,b), orange line).\\
\begin{figure}[h]
\centering
\includegraphics[scale=1.5]{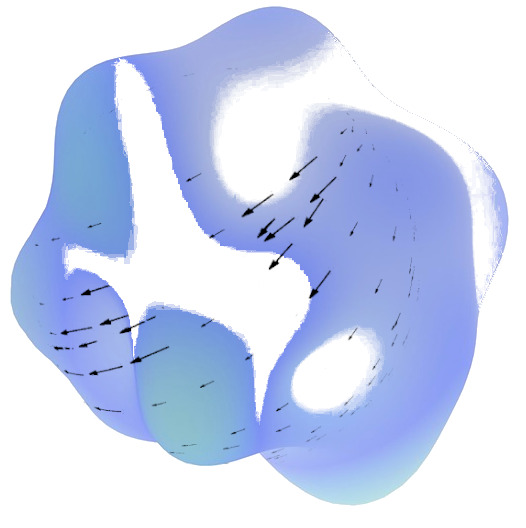}
\caption{\ansA{Shown is  a sketch of the reference phase space (blue blob), vectors $\{\mathbf{v}_{i\in[1,N]}\}$ (black arrows) pointing out from points  $\{\phi_{i\in[1,N]}\}$ and voids (white spots). Voids
are parts of the reference phase space which lack data (typically, due to short reference records or unreliable observations).
}}
\label{fig:ref_space3}
\end{figure}
}

{\bf Stochastic case iv} ($\mathbf{A}\ne0$ and $\mathbf{G}\ne0$;  criterion~\eqref{eq:opt2}; $S=2$). 
The stochastic correction results in even more accurate energy representation (figures~\ref{fig:mean_energy_2}(a,b), orange line).
Although the flow dynamics does not show qualitative differences compared to the deterministic case (iii) (figure~\ref{fig:mean_energy_2}(e)), \ansA{injection of noise into the advection velocity offers a smoother transition through voids compared to the deterministic case (iii)}. All in all, the proposed hybrid approach augmented with a stochastic correction of advection velocity and scale-aware energy control through the compensating forcing improves the intra- and inter-scale energy transfers towards the reference ones thus leading to the low-resolution flow dynamics that reproduces resolved flow features of the reference solution. 
Moreover, the hybrid approach also respects the regional stability property by keeping the hybrid solution in the vicinity of the reference phase space.

{\bf Stochastic ensemble simulations}. In this section we focus on stochastic ensembles to compare prediction skills 
of the hybrid model with its classical GFD counterpart. 
We consider a 50-member ensemble and run it over a two-month period.
We choose this setup as it is similar to those used by the European Centre for Medium-Range Weather Forecasts (ECMWF) for extended-range forecasts. In order to avoid an influence of the reference data, we compare the ensemble over the second validation year (where the reference data is unavailable), and also present the results over the reference year (where the reference data is available) for comparison.
\ansA{Along with the 50-member, 2-month long ensemble, we also present the results for a smaller ensemble (5 members), but computed over the
whole validation year.}

Every ensemble member of the hybrid model~\eqref{eq:pve_hybrid} starts from a perturbed reference initial condition. The \ansA{low-resolution} QG model~\eqref{eq:pve} starts from its own initial condition after the spinup, because starting it from the reference initial condition is not allowed as it might give the model an advantage over the drift time (the time the model drifts away from the reference energy manifold
until it is equilibrated onto its own lower-energy manifold). It takes around 5 years to equilibrate (not shown). The hybrid model has no drift (or more accurately, its drift is corrected), as its evolution is bounded to the reference energy manifold, and the ensemble start from perturbed reference conditions is therefore justified. However, if the hybrid model starts from an initial condition used for the \ansA{low-resolution} QG model, then it takes approximately one month for the solution to drift from the low-energy manifold of the \ansA{low-resolution} QG model to the high-energy reference manifold (not shown). This drift can be significantly accelerated by increasing the nudging strength. 

\ansA{To proceed, we would like to draw the reader attention to the following notations:
the $q_i$ (normal font) refers to $\mathbf{q}$ (bold font) at the i-th layer, while $\mathbf{q}_i$ refers to either
the i-th reference record or to the i-th ensemble member. There should be no confusion, as we use local indices for both reference records and ensemble members.}

\begin{figure}[h]
\hspace*{-0.5cm}
\begin{tabular}{ll}
\hspace*{0.4cm}(a)  & \hspace*{0.4cm}(b) \\
\begin{minipage}{0.5\textwidth}\includegraphics[scale=0.13]{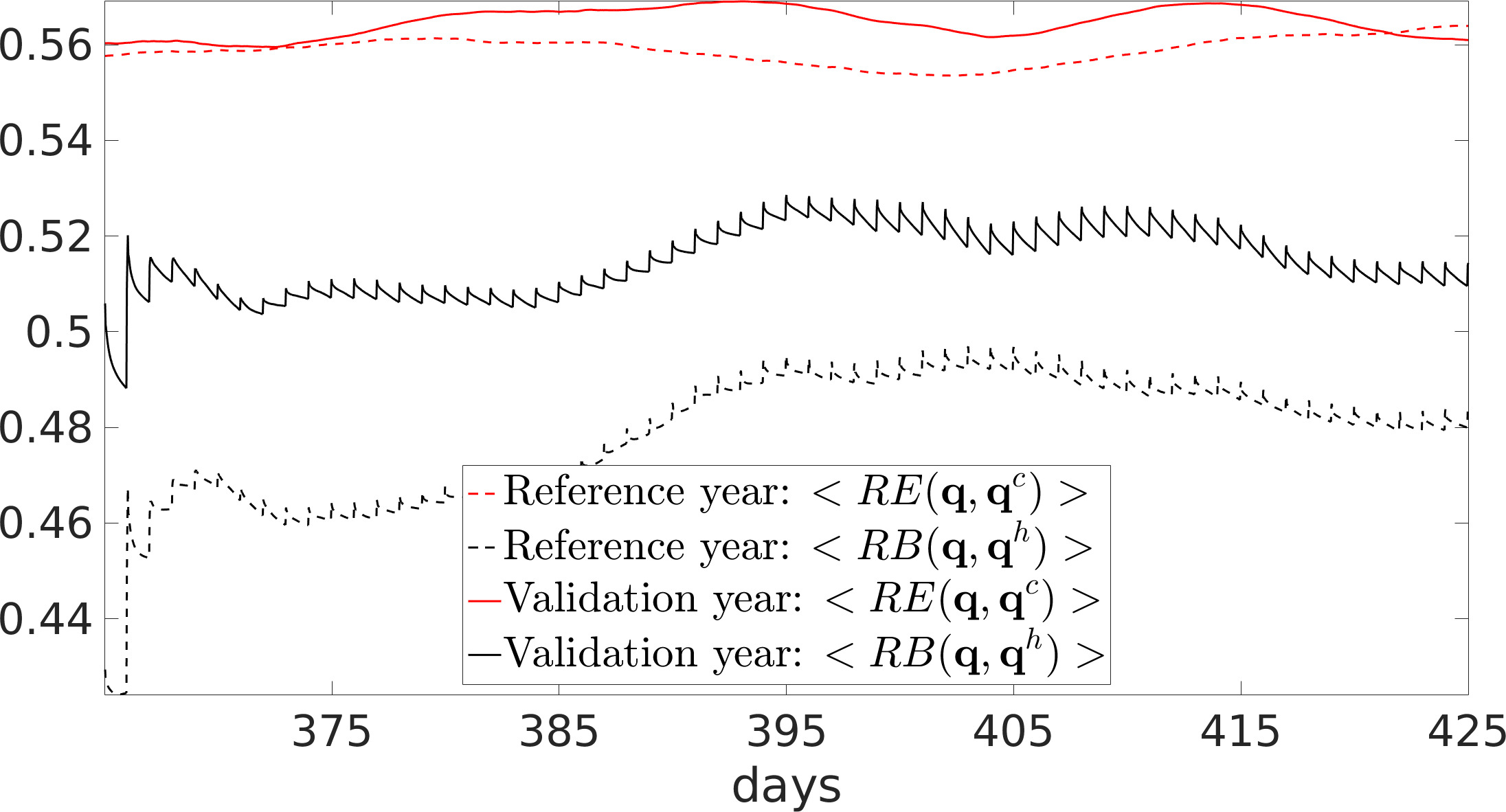}\end{minipage} &
\begin{minipage}{0.5\textwidth}\includegraphics[scale=0.13]{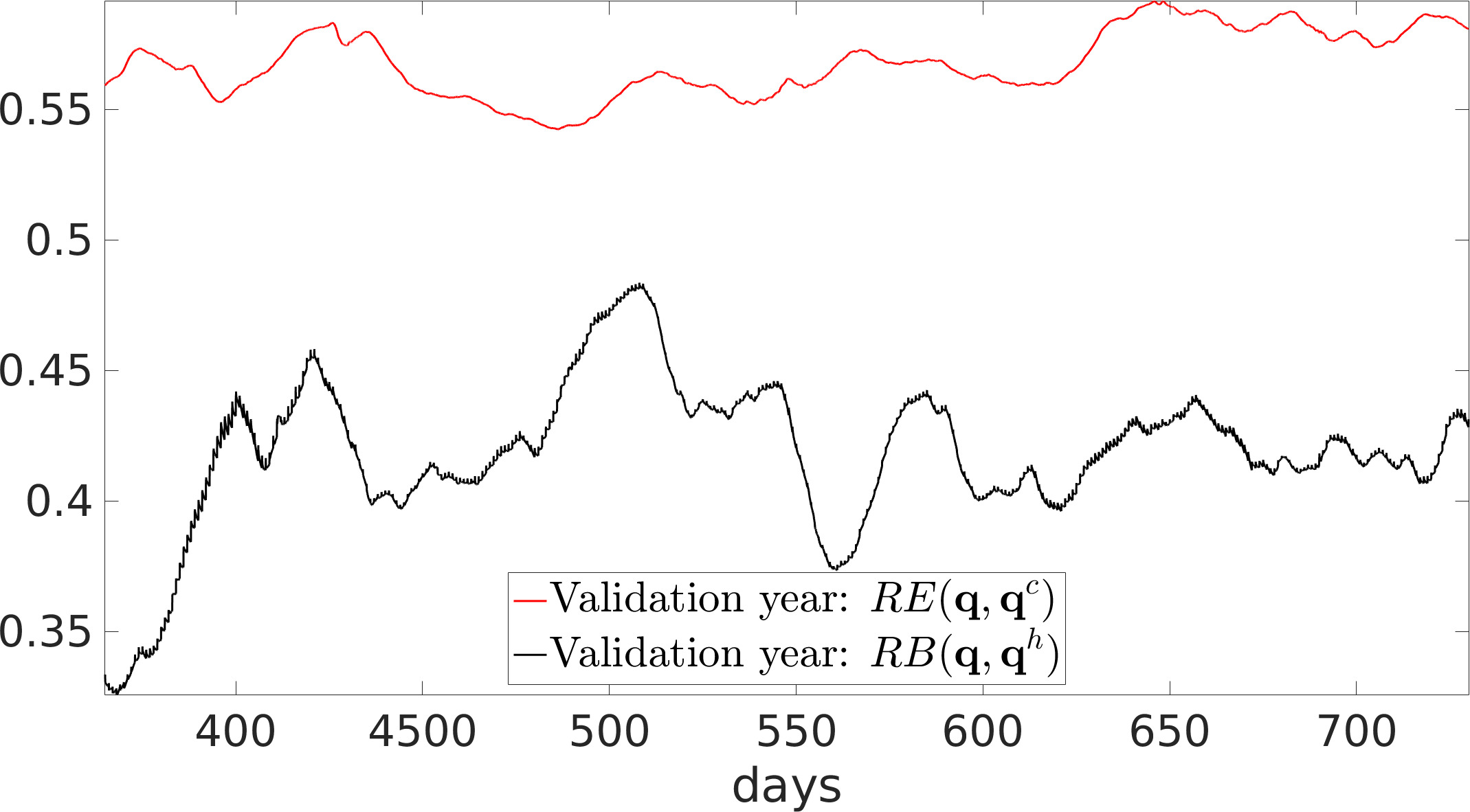}\end{minipage}
\end{tabular}
\caption{\ansA{Shown is (a) $\langle RB \rangle$ (black) and  $\langle RE \rangle$ (red), and (b) $RB$ (black) and  $RE$ (red).}
The results are presented for the reference year (dashed line) over which the reference data is available and 
for the validation year (solid line) over which the reference data is not available.
The results clearly show that the hybrid ensemble forecast is more accurate than that of the \ansA{low-resolution} QG model over the validation year.
The results for the reference year are given for comparison.}
\label{fig:re}
\end{figure}

The results are presented in figure~\ref{fig:re} in terms of relative bias (RB):\\
\ansA{
\begin{equation}
RB(\mathbf{q},\mathbf{q}^h):=\frac{\|\mathbf{q}-\overline{\mathbf{q}}^h\|_2}{\|\mathbf{q}\|_2},\quad 
\overline{\mathbf{q}}^h:=\frac{1}{N}\sum\nolimits^N_{i=1}\mathbf{q}^h_i\,,
\label{eq:rb1}
\end{equation}
$l_2$-norm relative error (RE):\\
\begin{equation}
RE(\mathbf{q},\mathbf{q}^c):=\frac{\|\mathbf{q}-\mathbf{q}^c\|_2}{\|\mathbf{q}\|_2},
\label{eq:rre2}
\end{equation}
as well as a 2-month $\langle RB \rangle$ and $\langle RE \rangle$ (which are RB and RE averaged over the validation year):\\
\begin{equation}
\begin{array}{l}
\displaystyle
\langle RB(\mathbf{q},\mathbf{q}^h)\rangle :=\frac16\sum\nolimits^5_{i=0} RB(\mathbf{q}_{[2i,2(i+1)]},\mathbf{q}^h_{[0,2]})\,,\\
\\
\displaystyle
\langle RE(\mathbf{q},\mathbf{q}^h)\rangle :=\frac16\sum\nolimits^5_{i=0} RE(\mathbf{q}_{[2i,2(i+1)]},\mathbf{q}^h_{[0,2]})\,,\\
\end{array}
\label{eq:rb2}
\end{equation}
with $N=50$. Since we only have a 2-month ensemble run (denoted as $\mathbf{q}^h_{[0,2]}$ in~\eqref{eq:rb2}), we extract six 2-month long intervals (with a 2-month step) from the validation year (it is denoted as $\mathbf{q}_{[2i,2(i+1)]}$ for $i=[0,5]$ 
in~\eqref{eq:rb2}), compute the relative bias~\eqref{eq:rb1} and the relative error~\eqref{eq:rre2} for every interval, and then average up the result over the six intervals as shown in~\eqref{eq:rb2}. And, we do the same for the low-resolution QG solution $\mathbf{q}^c$. 
The averaged RB and RE show that the hybrid solution stays closer to the reference solution than the low-resolution QG solution.
This happens because the hybrid solution lies on the reference energy manifold, while the low-resolution QG solution does not.
This finding is also confirmed by the results presented in figure~\ref{fig:re}b, which shows RB and RE for a smaller ensemble (5 members),
but {\it computed over the whole validation year}.
}\\

As seen in figure~\ref{fig:re}a, the ensemble forecast of the hybrid is more accurate
than that of the \ansA{low-resolution} QG model~\eqref{eq:pve}. Moreover, this forecast is systematically better not only over two months, but over the whole validation year.
It demonstrates the potential of the hybrid model to produce reliable forecast well beyond the reference dataset. 
The short fluctuation of the relative bias over the first $\sim$24 hours into the simulation is due to the adjustment of the hybrid model to the reference energy manifold. It happens as the perturbation of the reference initial condition can be outside of the reference energy band. The saw-like behaviour of the relative bias is due to the correction of the hybrid solution towards the reference energy manifold.
\ansA{This correction does not introduce discontinuities into the flow fields,
because we optimize the energy amplitudes on the solution of the hybrid model. To put it another way, the optimized solution 
is a solution to the hybrid model which guarantees the necessary smoothness of the fields. The saw-tooth behaviour can be avoided
(or significantly inhibited) when using the optimization over shorter periods.}

\ansA{It is worth mentioning that the reference data is used any time it is needed to compute the $\mathbf{G}$ and $\mathbf{A}$, and therefore to keep the hybrid solution
on the reference energy manifold. If a future state lies on the reference energy manifold then 
the hybrid model can predict it, because the evolution of the hybrid model is bound to this manifold. 
However, if a future state is not on the reference energy manifold then the hybrid model cannot predict it.
In other words, the reference data set should be a representative sample of the flow dynamics.
Obviously, the reference energy manifold can be extended by widening the reference energy band so as to include future 
states that are not sampled.
}

\section{Conclusions and discussion\label{sec:conclusions}}
In this study we have proposed a hybrid approach based on deterministic and stochastic energy-aware hybrid models that combine the principle of regional stability used in the hyper-parameterization approach and classical~\ansA{quasi-geostrophic} model at low resolution that cannot reproduce high-resolution 
reference flow features which are, however, resolved at low resolution.
The hybrid approach utilizes the velocity correction and energy-aware forcing (compensating for the tendency of the classical model to leave the reference energy manifold)
in order to improve intra- and inter-scale energy transfers towards the reference ones through the control of energy on specific scales.

We have tested the hybrid approach on the three-layer QG model and studied how different forcing terms affect its energy 
and flow dynamics. We have found that the forcing based on the HP-like nudging and constant 
scale amplitudes does not improve the hybrid solution towards the reference one, as the hybrid solution energies are outside of the reference energy band.
The first optimization criterion improves the situation but the energy of the hybrid model becomes contaminated with high-frequencies (resulting from constant corrections 
of the trajectory towards the reference phase space) which are not present in the reference solution.
The forcing using the second optimization criterion significantly improves the hybrid solution so that it reproduces resolved on the coarse grid features of the reference flow. 
This improvement is achieved through keeping the kinetic and potential energies of the hybrid solution on the reference energy manifold and through respecting the regional stability.

In studying the deterministic and stochastic hybrid QG models we have found that the stochastic model provides a marginally more accurate representation of the energy compared to the deterministic 
model, while the flow dynamics is qualitatively the same for both models. However, the stochastic model
\ansA{offers a smoother transition through voids compared to the deterministic case}
and also allows to produce ensembles
which are systematically more accurate than the solution of the \ansA{low-resolution} deterministic QG model as well as 
to apply a data assimilation methodology based on ensemble particle filters which we will consider in a sequel to this paper. 

\ansA{The fundamental difference between the proposed approach and the general data assimilation (DA) methodology is that DA tries to force the model to follow the reference solution by adjusting 
the initial condition of the model or by nudging the model solution to the reference one based on partial observations of the reference.
In short, DA tries to match individual trajectories (the model solution and the reference solution).
The hybrid approach is fundamentally different: it tries to match phase spaces (the model phase space and the reference phase space) by manipulating the energy on different spatial scales. The reference phase space (the blue blob in figure~\ref{fig:ref_space}) changes much slower (as a geometrical object) than individual trajectories evolving on it.
In a sequel to this work we will apply DA to both the original and the hybrid model.
}

The main advantage of the hybrid approach is that it allows for any physics-driven flow recombination within the reference energy band and reproduces resolved reference flow features in low-resolution models outside of the reference sample.
Hybrid models offer appealing benefits and flexibility to GFD modellers and the forecasting community, as (1) they are computationally cheap (a hybrid run is on the scale of the coarse-grid model simulation); (2) can use both numerically-computed flows and observations; (3) require minimalistic interventions into to the dynamical core of the underlying model, only the advection velocity is corrected and 
one term is added to the right hand side; (4) produce more accurate ensemble forecasts than classical GFD models.

\ansA{It is worth noting that for the use of the hybrid approach in realistic simulations with strong interannual variations,
one should use longer reference records to make sure they are a representative sample of the reference solution.
One of the most important criterion is to sample the correct reference energy band so that the hybrid solution 
could evolve within it for a long time.}

The ability of energy-aware stochastic hybrid model to provide high-quality solutions 
(given the fact that the computational cost of hybrid runs is that of the low-resolution ocean model with modest overhead)
allows for the use of larger ensembles and longer forecast horizons compared with the reference runs. 
The hybrid approach offers a great resolution scalability and can be used to parametrise the effect of any high-resolution process into 
lower-resolution systems (e.g. mesoscales processes in larger-scale processes, sub-mesoscale processes in mesoscale-resolving, etc.), and therefore it can be readily adopted by forecasting systems to increase the resolution of the resolved dynamics.

We conclude that the hybrid approach has the potential to be used for long-term climate projections with~\ansA{primitive-equations} low-resolution GFD models, which are currently not capable of playing this role.

\section*{Open Research Section}
\noindent
Simulation data produced and analysed in this study can be found at\\
\noindent
https://doi.org/10.5281/zenodo.10722741 .
The reference QG solution and the coarse-grid runs are also available at
https://doi.org/10.5281/zenodo.10722741 .
The QG software is available on the HPC cluster of Imperial College London and can be accessed by contacting
the HPC Support at hpc-support@imperial.ac.uk .
The GALAHAD optimization library is available at https://www.galahad.rl.ac.uk/download.html~.


\acknowledgments
Igor Shevchenko thanks the Natural Environment Research Council for the support of this work through the projects CLASS and ATLANTIS (P11742).
Dan Crisan thanks the European Research Council (ERC) under the European Union’s
Horizon 2020 Research and Innovation Programme for the partial support of this work through ERC, Grant Agreement No 856408.
The authors thank unknown referees for their comments and suggestions which helped us improve the manuscript.

%
%

\bibliography{refs}

%
%
%
%
%

\end{document}